\documentclass[conference,compsoc]{IEEEtran}

\def\BibTeX{{\rm B\kern-.05em{\sc i\kern-.025em b}\kern-.08em
    T\kern-.1667em\lower.7ex\hbox{E}\kern-.125emX}}

\usepackage[compact]{titlesec}
\usepackage{algpseudocode}
\usepackage{algorithm}
\usepackage{amsmath}
\usepackage{amsfonts}
\usepackage{amssymb}
\usepackage{bm}
\usepackage{cite}
\usepackage[labelfont=bf, textfont=normal]{caption}
\usepackage{enumitem}
\usepackage[T1]{fontenc}
\usepackage{makecell}
\usepackage{mathtools}
\usepackage{multirow}
\usepackage{pifont}
\usepackage{subcaption}
\usepackage{xspace}
\usepackage{xcolor}
\usepackage{chngpage}
\usepackage{url}
\usepackage{hyperref}
\usepackage[nameinlink]{cleveref}
\usepackage{enumitem}

\renewcommand{\paragraph}[1]{\noindent\textbf{#1.}}

\makeatletter
\newcommand{\printfnsymbol}[1]{%
  \textsuperscript{\@fnsymbol{#1}}%
}
\makeatother

\newcommand{\System}{Specular\xspace} 
\newcommand{\Client}{SpecGeth}
\newcommand{\ClientErigon}{SpecErigon}
\newcommand{\RollupVM}{EVM\xspace} 

\newcommand{\GethProverLoC}{99\xspace} 
\newcommand{\ErigonProverLoC}{148\xspace}

\newcommand{\secprop}{security\xspace} 
\newcommand{\sota}{compiled\xspace} 
\newcommand{\thiswork}{native\xspace} 
\newcommand{\claim}{claim\xspace}
\newcommand{\claims}{claims\xspace}

\newif\ifcomments
\commentsfalse
\ifcomments
\newcommand{\todo}[1]{\textcolor{orange}{TODO: #1}}
\newcommand{\ujval}[1]{\textcolor{brown}{ujval: #1}}
\newcommand{\dawn}[1]{\textcolor{red}{dawn: #1}}
\newcommand{\zhe}[1]{\textcolor{teal}{zhe: #1}}
\newcommand{\franck}[1]{\textcolor{green!80!black}{franck: #1}}
\else
\newcommand{\todo}[1]{}
\newcommand{\ujval}[1]{}
\newcommand{\dawn}[1]{}
\newcommand{\zhe}[1]{}
\newcommand{\franck}[1]{}
\fi

\newif\ifnew
\newfalse
\ifnew
\newcommand{\new}[1]{\textcolor{blue}{#1}}
\else
\newcommand{\new}[1]{#1}
\fi

\newif\ifnewinternal
\newinternalfalse
\ifnewinternal
\newcommand{\newint}[1]{\textcolor{blue}{#1}}
\else
\newcommand{\newint}[1]{#1}
\fi

\newif\iflowprio
\lowpriofalse
\iflowprio
\newcommand{\todolowprio}[1]{\textcolor{orange}{todo (p1): #1}}
\else
\newcommand{\todolowprio}[1]{}
\fi

\newif\ifreducecaptionvspace
\reducecaptionvspacefalse
\ifreducecaptionvspace
\newcommand{\captionspace}{\vspace{-1em}}
\else
\newcommand{\captionspace}{}
\fi

\begin{document}

\title{Specular: Towards Secure, Trust-minimized Optimistic Blockchain Execution}

\newif\ifauthor
\authortrue

\ifauthor
\author{
\IEEEauthorblockN{
Zhe Ye\textsuperscript{$\dagger$\textsuperscript{\textsection}}, 
Ujval Misra\textsuperscript{$\dagger$\textsuperscript{\textsection}}, 
Jiajun Cheng\textsuperscript{$\ddagger$},
Wenyang Zhou\textsuperscript{$\mathparagraph$},
and Dawn Song\textsuperscript{$\dagger$} 
} %
\IEEEauthorblockA{
\textsuperscript{$\dagger$}University of California, Berkeley
}
\IEEEauthorblockA{
\textsuperscript{$\ddagger$}ShanghaiTech University
}
\IEEEauthorblockA{
\textsuperscript{$\mathparagraph$}University of Cambridge
}
}
\fi

\maketitle
\IEEEpeerreviewmaketitle

\ifauthor
\begingroup\renewcommand\thefootnote{\textsection}
\footnotetext{Equal contribution.}
\endgroup

\fi

\begin{abstract}
An optimistic rollup (ORU) scales a blockchain's throughput by delegating computation to an untrusted remote chain (L2), refereeing any state claim disagreements between mutually distrusting L2 operators via an interactive dispute resolution protocol.
State-of-the-art ORUs employ a monolithic dispute resolution protocol that tightly couples an L1 referee with a \textit{specific L2 client binary}---oblivious to the system's higher-level semantics.
We argue that this approach (1) magnifies monoculture failure risk, by precluding trust-minimized and permissionless participation using operator-chosen client software;
(2) leads to an unnecessarily large and difficult-to-audit TCB; and, 
(3) suffers from a frequently-triggered, yet opaque upgrade process---both further increasing auditing overhead, and broadening the governance attack surface.

To address these concerns, we outline a methodology for designing a secure and resilient ORU with a minimal TCB, by facilitating opportunistic 1-of-N-version programming.
Due to its unique challenges and opportunities, we ground this work concretely in the context of the Ethereum ecosystem---where ORUs have gained significant traction.
Specifically, we design a semantically-aware proof system, \textit{natively targeting} the EVM and its instruction set.
We present an implementation in a new ORU, \textit{Specular}, that opportunistically leverages Ethereum's existing client diversity with minimal source modification, demonstrating our approach's feasibility.
\end{abstract}


\newcommand{\ACCEPT}{\texttt{ACCEPT}\xspace}
\newcommand{\REJECT}{\texttt{REJECT}\xspace}

\newcommand{\swapCmd}{\texttt{SWAP}\xspace}
\newcommand{\dupCmd}{\texttt{DUP}\xspace}

\newcommand{\worldState}{\bm{\sigma}\xspace}
\newcommand{\machineState}{\bm{\mu}\xspace}
\newcommand{\codeState}{\bm{\kappa}\xspace}

\newcommand{\sInit}{\omega\xspace}
\newcommand{\sFinal}{\omega'\xspace}
\newcommand{\sInter}{\omega_{\texttt{int}}\xspace}
\newcommand{\sBlock}{\omega_{\texttt{b}}\xspace}
\newcommand{\hInit}{h\xspace}
\newcommand{\hFinal}{h'\xspace}

\newcommand{\data}{d\xspace}

\newcommand{\HStack}{H_{\texttt{stack}}\xspace}
\newcommand{\HLogs}{H_{\texttt{logs}}\xspace}
\newcommand{\HLogsBloom}{H_{\texttt{logs+bloom}}\xspace}
\newcommand{\HOSS}{H_{\texttt{OSS}}\xspace}
\newcommand{\KECCAK}{\texttt{KEC}\xspace}

\newcommand{\accNonce}{\texttt{nonce}\xspace}
\newcommand{\accBalance}{\texttt{balance}\xspace}
\newcommand{\storageRoot}{\texttt{storageRoot}\xspace}
\newcommand{\codeHash}{\texttt{codeHash}\xspace}
\newcommand{\codeSize}{\texttt{codeSize}\xspace}
\newcommand{\codeRoot}{\texttt{codeRoot}\xspace}

\newcommand{\offset}{\texttt{offset}\xspace}
\newcommand{\opcode}{\texttt{opcode}\xspace}
\newcommand{\op}{\texttt{op}\xspace}

\newcommand{\stackMS}{\machineState_{\bm{s}}\xspace}
\newcommand{\stackMSPrime}{\machineState'_{\bm{s}}\xspace}
\newcommand{\sPopped}{p\xspace}
\newcommand{\sPushed}{q\xspace}

\newcommand{\memMS}{\machineState_{\bm{m}}\xspace}
\newcommand{\pcMS}{\machineState_{\texttt{pc}}\xspace}
\newcommand{\returnMS}{\machineState_{\bm{o}}\xspace}

\newcommand{\bytecodeEE}{I_{\bm{b}}\xspace}
\newcommand{\instructionEE}{\bytecodeEE[\pcMS]\xspace}
\newcommand{\substateLog}{A_{\bm{l}}\xspace}

\newcommand{\KNVP}{KNVP\xspace}

\section{Introduction}
\interfootnotelinepenalty=10000

\begin{figure}[t]
  \centering
  \includegraphics[width=0.98\linewidth]{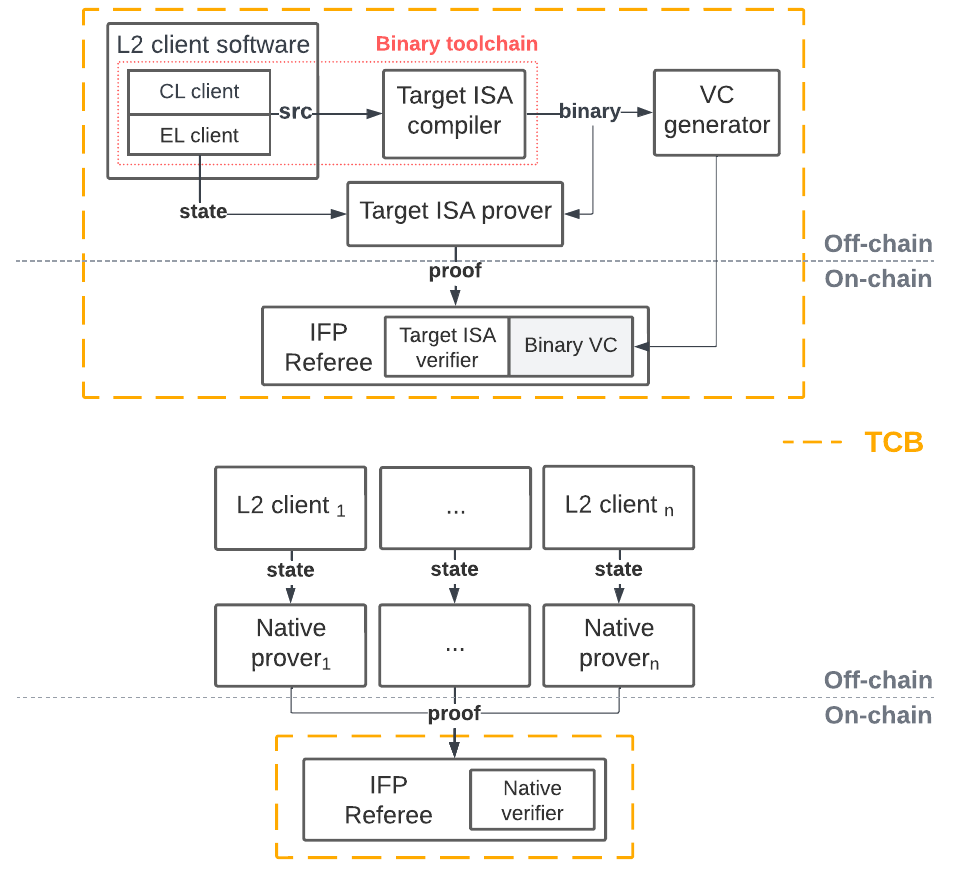}
  \caption{
  \textbf{Proof system toolchain comparison.}
  (Top) state-of-the-art ORUs suffer from a large and complex TCB. 
  The binary toolchain of the \sota approach is inherently resistant to TCB simplification.
  (Bottom) the \thiswork approach, instantiated with $1$-NVP, enables a simpler TCB.
  }
  \captionspace{}
  \label{fig:tcb}
\end{figure}

Public blockchains, such as Ethereum \cite{wood2014ethereum}, have struggled to scale with growing demand, resulting in exorbitant transaction fees during congestion. 
Blockchain nodes are generally required to disseminate, reach consensus over and execute transactions---presenting multiple potential performance bottlenecks.
A recent line of work \cite{kalodner2018arbitrum, mccorry2021sok, yee2022shades, bousfield2022arbitrum, optimism-bedrock} aims to mitigate the execution bottleneck, by securely offloading computation to a more powerful off-chain system (L2) operated by untrusted parties (\textit{validators}), and efficiently guaranteeing state transition validity on-chain (L1).
\textit{Optimistic rollups} (ORUs) are currently the most popular of such protocols, with several billions of dollars in total value locked, despite outstanding security risks \cite{l2beat}.

\newcommand{\T}{\mathbb{T}\xspace}
\newcommand{\SM}{\mathbb{S}\xspace}
\newcommand{\IFP}{\mathbb{D}\xspace} 

An ORU protocol defines, at minimum, 
(1) an off-chain execution semantics, and 
(2) an on-chain bridge \cite{mccorry2021sok}, secured by a dispute resolution protocol that enforces the execution semantics. \ujval{tighten}
During normal operation, an L2 validator locally applies the specified state-transition function $\T$ to inputs ordered and disseminated on L1. 
The validator then makes claims about L2 state changes by posting state commitments to the bridge.
The finalization of a claimed state is delayed until its confirmation period has elapsed, allowing other validators to contest it if they disagree. 
\new{If a \claim is contested with a deviating counterclaim, an L1 referee must determine which is valid.
To do so, it engages participating validators in a dispute resolution protocol, consisting of an interactive fraud proof (IFP) game.
The protocol enables the referee to efficiently find the first step in a trace at which participants' claims deviate.
An honest validator may then submit a \textit{one-step proof} to convince the referee of the correct step-level transition.
Through this mechanism, false claims are ultimately rejected (within a delay).}
This ensures the security of the bridge under a \textit{1-of-n honest minority} assumption \cite{canetti2013refereed}.

An honest validator must by definition operate using L2 client software that conforms to the specification. 
As a matter of implementation convenience, state-of-the-art ORUs, such as Arbitrum Nitro \cite{bousfield2022arbitrum} and Optimism's Bedrock release \cite{optimism-bedrock}, rely on an IFP that referees disputes over the execution trace of---not the abstract specification semantics, but rather---\textit{a specific compiled binary} (from the client source), entrusted to conform to the specification.
That is, the referee neither verifies nor is necessarily aware of any higher-level semantics.
By conflating the binary with the protocol, the referee blindly enforces excessively over-constrained semantics at best, and incorrect exploitable semantics at worst. 
This approach therefore has three fundamental disadvantages: it 
(1) magnifies monoculture failure risk by hindering modular, trust-minimized and permissionless N-version programming (NVP); 
(2) leads to an unnecessarily large and complex trusted computing base (TCB) that is difficult to independently audit and impractical to formally verify; and, 
(3) suffers from a frequently triggered upgrade process, both increasing security audit overhead and broadening the governance attack surface.



First, by binding the verifier to a specific L2 client program binary, the protocol increases the risk of monoculture failures and weakens the credibility of the 1-of-n honest minority assumption. 
This is because validators are given limited operational choice and are required to run the pre-authorized binary during IFP execution. 
Consequently, invalid state transitions induced by software bugs can slip by undisputed---ultimately resulting in loss of user funds.
Ongoing community efforts \cite{optimism-decentralization, magi} to address this concern by introducing limited redundancy through \textit{in-protocol, monolithic} NVP \cite{chen1978n, knight1986experimental, breidenbach2018enter}---as we will establish in this work---face fundamental trust, permissioning and operational limitations, among others.

Second, because the L1 verifier is tasked with verifying the execution of the client at the target ISA instruction-level,
the TCB includes the source code of the L2 client software, target ISA compiler and vector commitment generator (as illustrated in \Cref{fig:tcb}).
Adding another layer of complexity to the TCB in practice, the refereed binary differs---as a function of both the source and target---from that of the validator's normal operation. 
Namely, the source is a \textit{modified} L2 client program that replaces select software components with oracles to abstract away I/O and non-determinism; the target is a different easily-emulated (and possibly custom \cite{bousfield2022arbitrum}) virtual ISA.
Altogether, this leads to a large attack surface and commensurately increases auditing overheads. 
Formal verification against an executable protocol specification is also infeasible in this regime, given the unbounded and concurrent nature of these programs.

Third, auditing overheads compound with the high frequency in upgrades resulting from a large TCB. 
Client programs, for example, are naturally upgraded more frequently than their underlying specification \cite{geth-github, eips}, in order to ship performance improvements and vulnerability patches.
Moreover, \textit{any} off-chain component upgrade requires an on-chain upgrade of the binary commitment in-tandem, putting L1 governance on the critical path.
Such upgrades not only require developer trust, but are also often slow, controversial and disruptive to the system \cite{kiayias2022sok}.
For this reason, ORU protocol maintainers have expressed a desire to forfeit upgrade control over the L1 referee in the long-term \cite{optimism-decentralization}. 
However, this is ill-advised and unrealistic, given \textit{all} off-chain software upgrades---including vulnerability patches---must be reflected on-chain in a new binary commitment.


\begin{table*}[t]
  \centering {\small
  \begin{tabular}{|c|c|c|c|c|c|c|}
    \hline
    \multirow{3}{*}[0.5em]{\textbf{ORU system}} & 
    \multicolumn{6}{c|}{\textbf{Properties}} \\
    \cline{2-7}
    &
    \makecell{Exec target} &
    \makecell{IFP target} &
    \makecell{NVP capability} &
    \makecell{TCB} &
    \makecell{L1 upgrade trigger} &
    \makecell{Efficient referee} \\
    
    \hline
    
    Fuel v2 \cite{fuel-github} & 
    FuelVM &
    FuelVM \cite{fuelv2-specs} &
    1-of-N$^{\ *,a}$ &
    L1 referee$^*$ &
    L2 spec upgrades &
    \checkmark \\
    
    \hline
    Arbitrum v1 \cite{kalodner2018arbitrum} & 
    AVM &
    AVM \cite{kalodner2018arbitrum} & 
    1-of-N$^{\ *,a}$ & 
    L1 referee$^*$ &
    L2 spec upgrades &
    \checkmark \\
    
    \hline
    Arbitrum Nitro$^{\dag}$ \cite{bousfield2022arbitrum} & 
    \new{EVM+} &
    WASM \cite{wasm} &
    \new{K-of-N$^{\ *,p,m}$} &
    Full system &
    L2 software upgrades &
    \checkmark \\
    
    \hline
    Optimism Bedrock$^{\ddag}$ \cite{optimism-bedrock} & 
    EVM &
    MIPS \cite{hennessy82mips} &
    K-of-N$^{\ *,p,m}$ &
    Full system &
    L2 software upgrades &
    \ \checkmark$^*$ \\
    
    \hline
    \textit{\System} & 
    EVM &
    EVM \cite{wood2014ethereum} &
    1-of-N &
    L1 referee &
    L2 spec upgrades &
    \checkmark \\
    
    \hline
  \end{tabular}}
  \caption{
  \textbf{System comparison.} 
  $^*$ denotes \textit{theoretically} realizable properties (applying a liberal interpretation), with varying feasibility.
  $^{\dag}$ 
  \new{targets the EVM with extended semantics, making opportunistic NVP less feasible}; 
  $^{\ddag}$ has a nascent but under-specified proposal \cite{optimism-decentralization}.
  $^a$ denotes ad hoc-only NVP capability;
  $^p$ denotes permissioning limitations;
  $^m$ denotes monolithism.
  }
  \captionspace{}
  \label{table:sys-comparison}
\end{table*}

\paragraph{Requirements} 
An ORU should therefore fulfill four high-level design goals.
First, the protocol's IFP should \textit{at minimum}, permit NVP, and \textit{preferably} facilitate it in a modular, trust-minimizing and permissionless form, to achieve resiliency against bugs.  
Second, the TCB must be sufficiently small and simple to enable effective security audits and ideally, resiliency through NNVP \cite{breidenbach2018enter} and/or formal verification against an executable specification \cite{hildenbrandt2018kevm, cassez2023formal}. 
Third, TCB upgrades should ideally be only as frequent as upgrades to the protocol specification; semantically-equivalent L2 client software upgrades should not be hindered by L1 governance. 
Last, these properties should not come at an adverse cost to performance during normal operation and dispute resolution.

\begin{figure}[t]
  \centering
  \includegraphics[clip, trim=0cm 0.5cm 0cm 0.9cm, width=0.75\linewidth]{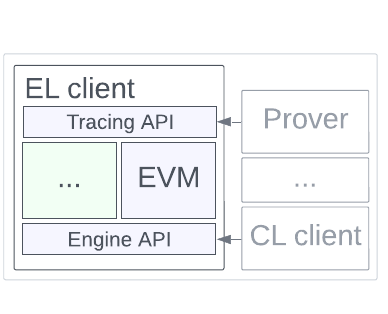}
  \caption{
  \textbf{Opportunistic 1-NVP in \System.} 
  \System opportunistically reuses an Ethereum EL client for its own execution-layer. (Purple) Components are minimally modified to enable finer-grained control over payload creation, and expose sufficient trace state to the prover.
  (Green) All other components are fully reused. \ujval{update}
  }
  \captionspace{}
  \label{fig:overview}
\end{figure}

\paragraph{Native IFPs}
We argue that compiled binary-based IFPs are vulnerability-prone.
We instead propose an alternative \textit{L2-native} IFP protocol design approach that proves and verifies L2 semantics \textit{directly}. 
That is, the proof system should make minimal assumptions beyond the specified semantics.
This unlocks several advantages relative to prior work. 
Most notably, different L2 client software systems share an abstract higher-level execution trace, independent of their implementation details, and are therefore interoperable. 
That is, two validators running different software can still engage with one another in an IFP game.
As a result, only a single L2 client software system must conform to the specification to maintain ORU security.
We term this variant of traditional (majority-rule based) NVP as $1$-of-$N$ version programming, or $1$-NVP.

This approach permits out-of-protocol, \textit{permissionless} NVP, allowing any software implementing the specification to participate, without governance intervention \cite{optimism-governance}. 
Additionally, richer IFP protocol semantics provide modular, finer-grained control over redundancy at various abstraction levels.
For example, NVP can be applied to the L1 referee to achieve a more resilient and trustworthy TCB---\textit{without} adding complexity at L2 (as with the compiled approach).

Earlier generation ORUs \cite{kalodner2018arbitrum, fuel-github}, as outlined in \Cref{table:sys-comparison}, target the ISA of a \textit{bespoke VM} both in normal operation and at the IFP-level. Such VMs are natively designed to support IFPs efficiently.
Unfortunately, while they permit trust-minimized and permissionless $1$-NVP in theory, realizing this property in practice requires ad hoc, de novo development of independent client software systems.
Knight \& Leveson \cite{knight1986experimental} argue that the resiliency improvement from doing so does not merit the engineering effort required, due to its ineffectiveness in preventing correlated bugs.

However, we observe that this challenge can be circumvented.
There are existing efforts towards achieving client diversity at L1, motivated by a similar objective of reducing the likelihood of consensus failures \cite{dankradfeist23majorityclient}.
This provides a basis for \textit{opportunistically} bootstrapping multiple trustworthy L2 client software systems from existing L1 infrastructure.

\paragraph{Opportunistic $\textbf{1}$-NVP}
The Ethereum blockchain itself, as well as its execution semantics \cite{wood2014ethereum} is overwhelmingly popular.
The chain has demonstrated considerable resilience in spite of occasional mass client failures caused by software bugs \cite{galano20infura, geth-bug-aug21, yang2021finding}, owing to the diversity of permissionlessly participating L1 client software\footnote{In the cited case, the deciding factor was software \textit{version} diversity.}.
Its consensus rules uniquely incentivize client diversity \cite{dankradfeist23majorityclient}, and as a fortunate byproduct, has made client software available as an opportunistically reusable resource \cite{rodrigues2001base, castro2003base}. 
In this work, we therefore take particular interest in designing an L2-native IFP, specifically for Ethereum and its VM semantics.



Indeed, state-of-the-art ORU projects \cite{bousfield2022arbitrum, optimism-bedrock} have adapted one particular Ethereum execution-layer (EL) client, Geth \cite{geth-github}, to support L2 operation.
A validator runs this L2-adapted EL client, along with a custom consensus-layer\footnote{In a rollup, consensus is generally \textit{derived} deterministically from L1.} (CL) client.
The L1 referee enforces the semantics of a binary, generated by monolithically compiling a program containing cherry-picked components from both clients, to a lower-level target ISA.
Unfortunately, while such systems may still leverage a limited form of \textit{opportunistic NVP}, the IFP constrains the design space for doing so (see \Cref{table:sys-comparison}).

\textit{Combining opportunistic NVP with an L2-native IFP approach addresses these shortcomings.}
Specifically, we propose a one-step proof scheme that \textit{directly} targets a higher-level L2 semantics composed from that of Ethereum. 
The key challenge in this setting is to support proof generation directly over all EVM instructions, block creation and inter-transaction semantics, without significantly modifying L1 client internals. 
This facilitates $1$-NVP by preserving the ability to use any existing or future Ethereum client software that conforms to the specification.
Conventional wisdom in the blockchain industry \cite{optimism-evm-equivalence, qasession22} has held the sentiment that to do so would pose a significant challenge, due to the general complexity of the EVM.
\new{However, we demonstrate its feasibility, with the use of a simple authenticated data structure (ADS) \cite{tamassia2003authenticated} using standard primitives}---supporting efficient on-chain emulation of the EVM stack, memory, persistent storage and other auxiliary data structures.

\paragraph{Contributions}
To our knowledge, this is the first work to study the relationship between NVP techniques \cite{chen1978n, knight1986experimental, rodrigues2001base, castro2003base, breidenbach2018enter} and RDoC-based protocols \cite{canetti2011practical, canetti2013refereed, kalodner2018arbitrum, optimism-bedrock}.
We make both conceptual and technical contributions.
In \Cref{sec:framework}, we propose the use of an \textit{L2-native} IFP that enables opportunistic $1$-NVP---drawing from classic ideas in systems literature \cite{chen1978n, knight1986experimental, rodrigues2001base, castro2003base}, and more recent efforts in the blockchain setting \cite{breidenbach2018enter, dankradfeist23majorityclient}. 
We motivate the necessity of this approach in the context of prior work.

The rest of this work focuses on Ethereum specifically.
In \Cref{sec:evmnative}, we provide the first concrete scheme for an L2-native IFP to target the EVM.
Finally, we introduce \textit{\System}, a new secure and trust-minimized optimistic rollup.
\System leverages multiple Ethereum EL client implementations, namely Geth \cite{geth-github} and Erigon \cite{erigon-github}, adapted to support an L2-native IFP with only \GethProverLoC and \ErigonProverLoC lines-of-code modified respectively.




\section{Background}
We briefly survey relevant background in both traditional distributed systems and blockchain systems.
We also describe concurrent open-source efforts pursuing the application of NVP in ORUs \cite{optimism-decentralization}.
\label{sec:background}
\subsection{Refereed delegation}
\label{sec:refereed}
Refereed delegation of computation (RDoC) \cite{canetti2011practical, canetti2013refereed} consists of a family of protocols that allow a resource-bound client to efficiently and verifiably compute a function by delegating it to multiple untrusted servers, provided at least a single server is honest. 
Canetti et al. introduce an interactive protocol, instantiated from any collision-resistant hash function---summarized as follows.

Suppose a client delegates the computation of a function to two non-colluding servers.
If the servers unanimously agree on a result, the client accepts it immediately---since by assumption, one server is honest.
If the servers disagree, it initiates a bisection protocol with a logarithmic number of rounds to search for inconsistencies between the trace intermediate states of the servers' delegated computation.
In each round, the servers send the client binding commitments of their respective intermediate states at the computation step requested, generated using the hash function (to avoid sending the entire state). \ujval{update}

On identifying the inconsistency at the level of a single trace step (e.g. an instruction), the client determines which party is dishonest by 
(1) requesting the initial state, revealing the commitment previously received and agreed upon by both servers; and 
(2) locally emulating the step, accepting the result claimed by the honest server. 
Because the client is resource-bound, the protocol may employ the use of an authenticated data structure (ADS) \cite{tamassia2003authenticated, van2014versum, backes2013verifiable}, such as a Merkle tree \cite{merkle1987digital} or a more generic data structure \cite{miller2014authenticated, van2014versum}, to enable space-efficient emulation. 

In cases of unanimous agreement, there is no additional computational overhead for both the servers and the client.
During a disagreement, the overhead is poly-logarithmic in the size of the computation.
The protocol is \textit{computationally sound}, assuming a single server is honest.
It is also \textit{general-purpose} (supports any efficiently computable, deterministic function) and \textit{full-information} (requires no private state).

%

\subsection{Optimistic Rollups}
In this section, we describe a typical (but simplified) \textit{optimistic rollup} (ORU).
An ORU is specified by its off-chain execution semantics $\SM$
and a dispute resolution protocol $\IFP$ that defines state \claim confirmation and rejection semantics, to enforce $\SM$. 

\paragraph{Normal operation}
We define the following state machines, hierarchically (loosely following a similar model as \cite{wood2014ethereum}):
\begin{align*}
\SM_i &:= (\T_i, M_i, S) \ | \ 
\T_i : S \times M_i \rightarrow S \ | \ 1 \leq i \leq 4
\end{align*}
operating at the abstraction levels of (1) VM instructions, (2) transactions, (3) blocks, and (4) sequenced batches, respectively.
This captures the semantics at each key abstraction level of an L2 blockchain, for convenience---a sequenced batch contains multiple blocks; a block contains multiple transactions; and finally, a transaction involves the execution of multiple VM instructions.
Informally, for convenience we define each $\T_i$ to recursively subsume the semantics of $\T_j \ \forall j < i$, applied iteratively. 
For example, $\T_4$ applies a transaction batch to the L2 state in its entirety, building all contained transaction blocks. 

Users can submit their L2 transactions either directly to the L1 bridge, or as a matter of convenience and cost, to a uniquely-permissioned (and untrusted) L2 operator, known as a \textit{sequencer}.
We note that the role of the sequencer is orthogonal to this work, since an ORU can function securely without designating one; 
we therefore ignore its details for the rest of this work.

Validators---fulfilling the RDoC server role---read, decode and apply messages from the L1 bridge to their local state machine.
A validator submits claims to the bridge, attesting to a binding commitment on the state output by $\T_4$ (applied to the sequenced inputs), claiming the state's validity.
Once a claim is confirmed, the bridge unlocks any associated funds to be withdrawn.
However, the protocol delays its confirmation for a pre-determined time period\footnote{This is a system parameter, often set conservatively to 1 week.}, allowing any party to contest it by submitting a disagreeing counter-claim.
The claim is only confirmed at the end of this period if it is not rejected through a dispute. 

\paragraph{Dispute resolution}
An L1 referee allows any party to contest an unconfirmed \claim by submitting a disagreeing counter-claim.
This triggers the execution of a dispute resolution protocol between all parties that have attested to the \claims.
Ultimately, all but the correct \claim are rejected.




IFP protocols extend RDoC to the permissionless blockchain setting, where the L1 blockchain can be considered a trusted resource-bound client, and L2 validators the more computationally-powerful untrusted servers.
However, because there is no pre-determined committee of non-colluding servers that can be relied upon by the referee (as is assumed in RDoC protocols), ORUs typically aim to maximize participation (subject to computational and time constraints), by allowing \textit{any} party to participate as a validator (with dishonest behavior disincentivized through a financial penalty for attesting to rejected claims). 

As in the case of RDoC, \secprop is therefore guaranteed under a 1-of-$n$ honest-minority assumption.
That is, as long as a single honest party exists (and is live) to faithfully follow the protocol, an invalid claim will not be confirmed.

Kalodner et al. \cite{kalodner2018arbitrum} introduce a dispute resolution protocol, naturally extending the interactive protocol from \cite{canetti2011practical} to the permissionless blockchain setting---albeit with liveness limitations in the presence of multiple adversaries or sybils.
Recent work \cite{nehab2022permissionless, alvarez2023bold} addresses these limitations with more elaborate protocols, providing stronger liveness guarantees.
While these protocols differ---namely, in how they scale to multiple parties---all of them fundamentally rely on interactive trace bisection, punctuated with the verification of a \textit{one-step proof}.

\paragraph{One-step proof}
The first ORU systems \cite{kalodner2018arbitrum, fuelv2-specs} utilized straightforward constructions that natively targeted the off-chain execution semantics directly.
However, to target richer semantics (specifically, the EVM) more conveniently,
popular protocols \cite{bousfield2022arbitrum, optimism-bedrock} shifted towards constructions that target a lower-level ISA.
In this design, there is an initial offline setup phase (e.g. at protocol instantiation), where (1) a binary is compiled from an L2 client program (expected to implement the semantics correctly); 
(2) a vector commitment is generated from the binary (over each instruction); and 
(3) the commitment is submitted to the referee responsible for resolving disputes.

The target ISA is typically chosen to be a reduced instruction set (e.g. MIPS \cite{hennessy82mips}) to preserve proof system simplicity and support flexible compilation from higher-level languages.
We note that the ISA itself is often simplified further from its standard usage; for example, both Optimism (post-Bedrock) \cite{optimism-bedrock} and Arbitrum Nitro \cite{bousfield2022arbitrum} remove floating point arithmetic.

This work takes particular interest in the choice of proving target, and its implications on the security and trust assumptions of the broader system.

\subsection{N-version programming}
\label{sec:nvp}
N-version programming (NVP) \cite{chen1978n, knight1986experimental} is a classic systems technique, initially proposed to strengthen the \textit{fault tolerance}, or liveness, of software.
This is carried out first by preparing $N$ independently developed programs, intended to be functionally equivalent, and defining a decision function $f$ to transform the results of their execution on the same input.
In its simplest form, $f$ is invoked on the values output by the $N$ programs, selecting the value that appears in a majority quorum, if such a value exists.

\newint{In the absence of a quorum, a liveness-preferring $f$ may output a fallback value, while a safety-preferring $f$ would output abort $\bot$.
While NVP has traditionally been utilized in the former setting \cite{knight1986experimental}, we instead primarily consider its impact in the latter, for security \cite{breidenbach2018enter}.
In the rest of this work, we refer to the setup where the decision function outputs $\bot$ in the absence of a $K$-quorum (for $K > \frac{N}{2}$), as \KNVP.}

\subsubsection{NVP in distributed systems}
The effectiveness of NVP rests on the assumption that independently developed programs fail independently.
It has been shown, however, that this is often not the case in practice, due to hidden correlations \cite{knight1986experimental}.
Moreover, software development and maintenance costs grow linearly \cite{gray1991high}.

Another line of work \cite{rodrigues2001base, castro2003base, yin2003separating} partially addresses these concerns, particularly in the BFT-SMR setting, by seeking to exploit \textit{opportunistic} NVP.
In particular, BASE formally introduces the notion of reusing a set of distinct, existing off-the-shelf implementations---made equivalent via custom wrappers conforming to a \textit{common abstract specification}---instead of developing bespoke independent programs.
This eliminates the high development and maintenance costs of prior approaches. \\

\begin{figure}[t]
  \centering
  \includegraphics[width=\linewidth]{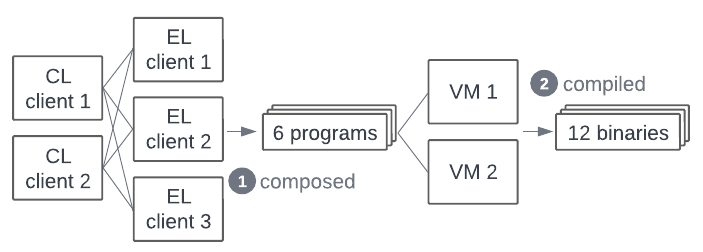}
  \caption{
  \textbf{Compilation-based \KNVP in ORUs.} 
  Components of independent CL and EL clients are composed as a single program.
  Programs are compiled to multiple targets.
  Each binary is committed to on-chain (not shown).}
  \captionspace{}
  \label{fig:knvp}
\end{figure}
\paragraph{NVP in blockchains}
The Ethereum network is the most prominent existing example of actively practiced NVP in the blockchain setting. 
As demonstrated for traditional BFT systems \cite{castro2003base, yin2003separating}, NVP prevents cascading failures and safety violations \cite{dankradfeist23majorityclient} in Ethereum---the effectiveness of which depends on the extent of client diversity across node operators \cite{clientdiversity}.
Ethereum full nodes run two types of client software---a consensus layer (CL) client, responsible for participating in consensus and validating block payloads received from the network;  
and an execution layer client (EL) client, responsible for constructing and executing blocks of transactions upon request from an authenticated CL client. 
Substantial community efforts have improved network resiliency through the independent development of multiple client implementations at both layers \cite{geth-github, erigon-github, reth-github, nethermind-github}.


\subsubsection{Opportunistic \KNVP in ORUs}
While client diversity exists in Ethereum, no rollup in production today employs NVP.
As discussed previously, this allows invalid state transitions induced by software bugs to slip by undisputed.
Optimism's protocol maintainers have acknowledged this as a security risk;
the community has therefore proposed---and is actively moving forward with the engineering of \cite{magi, asterisc}---a scheme that intends to mitigate the risks posed \cite{optimism-decentralization}.
We describe the intended scheme\footnote{A concrete spec. is nonexistent, but the scheme is straightforward.} and its design considerations below.

The scheme consists of a straightforward application of \KNVP within the compilation-based IFP paradigm, at the binary-level.
That is, the offline setup phase of the ORU involves committing to $N$ binaries rather than just 1.
Then, a disputed \claim is resolved by applying a $K$-quorum decision function to the outputs of $N$ independent invocations of the IFP game (one for each binary).
\newint{If no quorum exists, the decision function outputs $\bot$, indicating the chain must be halted until a more sophisticated referee---such as protocol governance---can manually resolve the dispute.
Assuming an implementation-honest party exists (i.e. conforms to the implementation, but not necessarily the specification), a submitted \claim is therefore confirmed only if at least $K$ binaries agree with it.}

\newint{Under this assumption, the following conditions are trivially derived.
First, the existence of a $K$-quorum for any given input is \textit{necessary} for liveness. 
Otherwise, dispute resolution is triggered when at least one binary deviates on an input, and the decision function---unable to establish a quorum---by definition outputs $\bot$.
Second, the system is secure if and only if for any input (1) a $K$-quorum does not exist, \textit{or} (2) the quorum output conforms to $\SM$.}

A key design choice in this paradigm is choosing which binaries to provision and commit to in-protocol. 
Since it is infeasible to develop several completely independent implementations, the Optimism proposal \cite{optimism-decentralization} suggests producing each binary through \textit{several combinations} of authorized software toolchain components, as illustrated in \Cref{fig:knvp}. 

This includes opportunistic reuse of Ethereum EL clients (with some non-trivial modifications to their semantics).
The source of both the CL and EL clients---normally maintained as separate programs---are partially composed to produce a single program, through white-box reuse of internal libraries.
The program is then compiled to multiple targets, to distribute trust among compilers and ISA targets.

We make two observations here.
First, the resulting binaries are assumed to be functionally equivalent to the corresponding systems utilized during normal execution that they represent, despite different sources and compilation targets.
Second, an exhaustive combination of all authorized toolchain components, as illustrated in the \Cref{fig:knvp}, is ideal to decorrelate bug incidence. 
However, this may not be feasible in practice due to performance impact and compatibility challenges, leading to a more selective binary provisioning.

\begin{figure*}
  \centering
  \includegraphics[width=\linewidth]{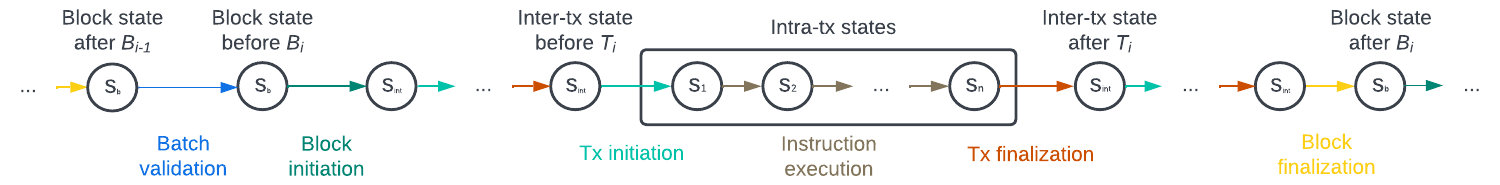}
  \caption{\textbf{Abstract L2 trace segment.} 
  \new{The L2 abstract trace includes state transitions between the following types of steps: (1) instruction execution, (2) transaction-level initiation \& finalization, (3) block-level initiation \& finalization, and (4) consensus-layer batch validation. 
  }}
  \captionspace{}
  \label{fig:state-transition}
\end{figure*}

Ultimately, while a \KNVP scheme provides incrementally improved defense against monoculture failures, it introduces other fundamental challenges. 
We enumerate these in the next section.
\section{An L2-native IFP}
\label{sec:framework}
In the same spirit as classic systems work which outlines a methodology for using abstraction in NVP to improve fault tolerance \cite{rodrigues2001base, castro2003base}, we propose the use of abstraction in NVP to improve security guarantees.
Namely, we argue for defining a common \textit{abstract trace} that decomposes the specified semantics $\SM$ to a sequence of \textit{steps}. 
Unlike in the \sota approach, steps are defined to preserve semantics at the highest-possible abstraction level, while preserving the tractability of efficient verification (without TCB bloat). 
The referee then operates over the abstract trace for a \claim, relying on a proof system that \textit{natively} targets the defined step-level semantics of $\SM$.

\Cref{fig:state-transition} illustrates how such a trace may be constructed, in a hierarchical bottom-up manner.
Specifically, a trace for all batches in a \claim is constructed such that 
(1) at the lowest level, each transaction consists of the sequence of executed L2 VM instructions---each considered a step---as specified by $\T_1$; 
(2) the remaining transaction-level ($\T_2$) semantics not captured are encapsulated in initiation and finalization steps for each transaction;
(3) the remaining block-level ($\T_3$) semantics not captured are encapsulated in block initiation and finalization steps for each block; and
(4) the remaining batch-level semantics ($\T_4$) not captured are encapsulated in a batch validation step for each batch.

We outline the trade-offs between the \sota and proposed approaches, the latter of which we refer to as the L2-native approach, below.
Specifically, we study their impact on the trust characteristics of the broader system.

\subsection{N-version programmability}

NVP is necessary to mitigate trust in any one particular implementation.
As previously touched upon, the two IFP approaches lead to fundamentally different NVP capabilities. 
The concrete instantiation of a \KNVP scheme in practice is constrained by trust, operational and access control considerations.

\subsubsection{In-protocol \KNVP}

\paragraph{TCB complexity}
The system is only secure if the \KNVP-transformed program conforms to the specification.
This implies that no $K$-quorum exists to produce semantically incorrect outputs.
A choice of an insufficiently large $K$ or $N$ binaries produced through highly-correlated toolchains weakens the credibility of this condition.
However, the precise configuration of these parameters presents trade-offs.

A cost-benefit analysis may indicate that the most sensible choice for $K$ is to favor safety over liveness, by setting $K:=N$ \cite{breidenbach2018enter}.
However, in trading off liveness, a slower higher-tier referee with a different (potentially weaker) trust model---such as protocol governance---bears larger decision-making responsibility.
The choice of $K$ can therefore have unclear trust implications.
Similarly, using a large set of binaries may only additively impact security if hidden correlations are absent; 
otherwise, the TCB is further bloated, while negatively biasing the decision function.


\paragraph{Operational impact}
Dispute resolution cost and operational overheads scale linearly with $N$, as 
the referee must conduct an IFP game for each binary.
Additionally, validators must themselves manage the operational complexity of deploying and executing each binary to participate in each game without additional external trust.
This presents a conflict, since uncorrelated bug incidence may require a large $N$. 
For example, exhaustive combination of authorized toolchain components, as previously described, results in a combinatorial explosion in the number of binaries.
The dispute resolution cost and operational complexity should therefore ideally grow independently of $N$.


\paragraph{Access control}
The use of a particular binary in dispute participation is \textit{permissioned}.
To enfranchise a binary in the decision function, it must be explicitly granted access to the dispute resolution protocol by committing to the binary on-chain.
This places another practical constraint on $N$: 
if the set of binaries is not fixed a priori, a trusted administrative mechanism (such as protocol governance) must exist to manually grant their addition and removal.
The burden on this mechanism compounds if toolchain upgrades are allowed, as we will elaborate further on in \Cref{sec:upgrades}.
Ideally, any binary that conforms to the specification can be utilized \textit{without permission}.



\subsubsection{Out-of-protocol 1-NVP}
The L2-\thiswork approach enshrines only the semantics $\SM$ in-protocol that the referee must enforce.
All L2 client software systems (i.e. the client and its prover) must interact against the same referee-enforced abstract trace.
An honest validator can therefore interoperably play against \textit{any other} validator in the system.
We call this $1$-NVP---\textit{this can be interpreted as a variant of NVP where the decision function is the referee itself.}

Critically, we assume that the referee's one-step proof verifier itself conforms to $\SM$.
The following condition is then trivially derived: 
the existence of a single honest party (conforming to $\SM$) is \textit{sufficient} for both security and liveness.
An honest party disputes invalid \claims to ensure security, and submits (and defends) valid \claims to ensure liveness---the latter guaranteed to the extent provided by the IFP game.

In this paradigm, there is no explicit access control mechanism---validators are afforded the option of running the L2 software system of their choice.
A validator can permissionlessly participate and win a dispute, utilizing \textit{any} L2 system that conforms to $\SM$.
Furthermore, the TCB, as we describe further in \Cref{sec:tcb}, is limited to only the referee.



\subsection{TCB trustworthiness}\label{sec:tcb}

The TCB of an ORU must be auditable, and amenable to further trust-minimization techniques, such as NNVP \cite{breidenbach2018enter} and formal verification, to ensure its trustworthiness.

\paragraph{Auditability}
In the \sota approach, because the referee enforces the execution of a binary at the target ISA instruction-level, inspecting the lower-level VM verifier alone is not sufficient to determine the enforced semantics. 
In practice, auditing the entire binary toolchain is necessary to determine with high confidence whether the semantics of the generated binary conform to the specification.
Moreover, as discussed in \Cref{sec:nvp}, in-protocol \KNVP necessitates audits for at least a majority of (if not all) the binary toolchains.
Such TCB bloat magnifies security vulnerability risk, and commensurately increases auditing overheads.


In contrast, because semantics are explicitly enforced by the verifier in the \thiswork approach, we consider a significantly reduced TCB size.  
With $1$-NVP, ORU security does not rely significantly on the correctness of any individual client program. 
The TCB of an L2-native ORU instantiated appropriately with permissionless participation through $1$-NVP, in the limit, only includes the referee.
This reduction of scope improves auditability, and renders the application of other trust-minimization techniques more tractable.

\paragraph{TCB size reduction}
We can further minimize the TCB through formal verification and NNVP. 
 
In the \sota approach, to verify the correctness of the binary against a formal specification, either its semantics must be verified directly, or through indirect verification of the individual toolchain components.
In practice, both paths are intractable, due to the size and complexity of each component.
While it \textit{is} practical to formally verify limited components, such as an on-chain RISC emulation-based verifier, this does not provide strong assurances regarding the enforcement of L2 semantics.
As described earlier, this is because a computationally sound proof system is still agnostic to binary-level bugs.
For the same reason, applying NNVP at the verifier-level is not generally useful.

On the other hand, in the L2-native approach, the VM verifier is the largest TCB component. 
It is therefore desirable to formally verify the VM verifier against a formal specification.
Fortunately, in the case of Ethereum, there are multiple specifications that can be reused---such as the Dafny-EVM \cite{cassez2023formal} and KEVM \cite{hildenbrandt2018kevm}.
By detecting inconsistencies between the L1 verifier implementation and the formal specification, the L1 verifier can be formally proven \textit{at the bytecode level} to conform to the intended L2 semantics (c.f. \Cref{sec:future}). 
This constrains the size of the TCB further to just the verification framework and specification.

Alternatively (or in composition), while we leverage $1$-NVP at L2, we can independently apply \KNVP to the VM verifier.
Unlike in the case of the \sota approach, \KNVP is applied to the verifier, rather than the binaries.
As a result, IFP interoperability between L2 systems does not break. 


\subsection{Upgrades}\label{sec:upgrades}

\paragraph{Frequency}
While blockchain client software is frequently upgraded (both for L1 and L2 networks), protocol specification upgrades tend to move slower. 
Ethereum, for example, has historically hard-forked at most twice each year \cite{eips}. 
Hard-forks that actually modify relevant semantics are generally even less common, at around once-a-year; 
semantic changes, such as those to consensus, often do not affect L2 execution semantics. 

Nevertheless, the \sota approach requires a bridge upgrade every time a binary toolchain component is upgraded, since changes to the source must be reflected in a new commitment on-chain.
For a protocol leveraging opportunistic \KNVP, this includes every time changes from an upstream client source are synced.
On the other hand, upgrades to the bridge of an L2-native ORU are as infrequent as those to the protocol specification.

\paragraph{Long-term access control}
A deliberately designed L1 \textit{protocol} will eventually stabilize, while \textit{implementations} will likely continue to commonly experience upgrades, e.g. to fix bugs, patch vulnerabilities and generally improve performance.
For the reasons highlighted above, the eventual forfeiture of bridge upgrade keys---a desired trust-minimization property \cite{optimism-decentralization}---is likely implausible. 

In comparison, the source code of L2-native ORU clients can stabilize in tandem with those of the L1 chain.
In the L2-native approach, the bridge must be upgraded only as frequently as semantic upgrades; therefore, the expectation is that with L1 stabilization, protocol governance can eventually forfeit upgrade keys, without forfeiting the ability to ship client-side changes.
We therefore argue that the safest and most practical path to relinquishing upgrade keys (and hence, mitigating trust in governance) is through an L2-\thiswork ORU design.

\paragraph{Transparency}
The size and complexity of the TCB in the \sota approach results in an opaque upgrade process, despite the use of opportunistic \KNVP. 
For example, EL client implementations are upgraded in a less transparent manner than the Ethereum specification, which undergoes a deliberate, public and peer-reviewed RFP process \cite{eips}. 
Additionally, while maintaining a fork of the upstream EL client---as is common for ORUs \cite{arb-nitro-github, optimism-bedrock}---allows developers to keep their L2 EL client in sync, it hinders the ability to distinguish what changes are semantically relevant to the L2 system.
Upgrades to other ad-hoc components, such as the compiler, are less transparent still.

In the native approach, there is a clear separation between verification of semantics and the client software systems implementing those semantics.
Therefore, it is easier to discern whether or not an upgrade can potentially affect the interpretation of semantics---auditors need only inspect the diff in the referee source.
\section{An EVM-native proof scheme}
\label{sec:evmnative}
In this section, we apply the takeaways from \Cref{sec:framework} to design a proof scheme with the trust implications of the broader system in mind.
Specifically, we target Ethereum semantics to take advantage of opportunistic $1$-NVP.
A one-step proof (OSP) convinces a verifier that given an initial EVM state---partially revealed and verified to be consistent against a commitment---executing the current instruction will result in a transition to the claimed final EVM state.
The scheme should address the following requirements. 
\begin{enumerate}[label=\arabic*.]
    \item
    \textbf{EVM-native.}
    The proof attests to the validity of a state transition at the granularity of a single EVM instruction (or inter-transaction operation). 
    All EVM instructions and inter-transaction operations are supported.
    \item
    \textbf{Specification-compatible.}
    A proof can be constructed exclusively from state specified by the EVM (i.e. without relying on knowledge of a specific EVM client implementation). 
    \item
    \textbf{Simple.}
    The scheme requires only standard cryptographic assumptions and can be achieved with a small, auditable TCB.
    \item
    \textbf{Efficient.}
    Proof size is at most logarithmic in the size of the EVM state $b$, linear in contract size and linear in bytes accessed.
\end{enumerate}
The key objective is therefore to design a \textit{simple} proof scheme for the EVM.
We also note that while proof verification efficiency is a concern, it is secondary to simplicity.
The IFP paradigm allows for tiered composition with a more efficient proof system (such as a SNARK-based construction \cite{pse}), \textit{without} inheriting stronger trust assumptions (we elaborate on this in \Cref{sec:future}).

The rest of this section describes the semantics of the system, followed by the associated one-step proof scheme. 
The construction uses standard assumptions, requiring only a collision-resistant hash function. 

\subsection{Preliminaries}
\label{sec:definitions}
We first summarize Ethereum's execution semantics. 
We build an L2 semantics upon this, followed by a description of our proof scheme.

\subsubsection{\new{L1 execution semantics}}
\label{sec:evm}
Ethereum is a permissionless, programmable blockchain that exposes a general-purpose state machine with a quasi-Turing complete ISA.
This subsection focuses on the EVM state relevant to transaction execution.
The EVM is a virtual stack machine that defines how bytecode instructions alter the Ethereum state. 
It has a volatile memory represented by a word-addressed byte array and a non-volatile storage, represented by a word-addressed word array.  
Both memory and storage are zero-initialized at all locations.
EVM bytecode is stored in virtual read-only memory, accessible only through a specialized instruction.
The EVM state (both volatile and non-volatile) is split across the EVM world state, machine state, accrued substate and environment information.
We summarize each of these below. 
A full definition of this state can be found in the Ethereum Yellow Paper \cite{wood2014ethereum}. 
The semantics and gas cost corresponding to each instruction which mutates the state can be found there in Appendix H.

The world state $\worldState$ is a mapping between addresses and account states, stored in a Merkle Patricia tree (trie).
Each account $\worldState[a]$, identified by its address $a$, is comprised of an intrinsic monetary $balance$ and transaction count $nonce$.
An account is also optionally associated with storage state and EVM code through a $\storageRoot$ (256-bit hash of storage MPT root) and $\codeHash$ (hash of bytecode stored in a separate state database) respectively.
All fields are mutable except $\codeHash$, which is write-once on contract creation.

The machine state $\machineState$ of the current messsage-call or contract creation is a 6-tuple $(g, pc, \bm{m}, i, \bm{s}, \bm{o})$ of the remaining gas available $g$, program counter $pc \in \mathbb{N}_{256}$, memory contents $\bm{m}$, active number of words in memory $i$, stack contents $\bm{s}$, and the return data from the previous call $\bm{o}$\footnote{This field is omitted from the definition of $\machineState$ in the EVM specification \cite{wood2014ethereum}---possibly unintentionally---but can be found in Appendix H where the semantics of \texttt{RETURNDATASIZE} and \texttt{RETURNDATACOPY} are described.}.

The execution environment information is a tuple of read-only data $I \coloneqq (I_a, I_o, I_p, I_{\bm{d}}, I_s, I_v, I_{\bm{b}}, I_H, I_e, I_w)$ that can be accessed by specific instructions. 
This includes executing bytecode's account address $I_a$, transaction sender address $I_o$, gas price $I_p$, input data $I_{\bm{d}}$, invoker account address $I_s$, monetary value $I_v$, bytecode $I_{\bm{b}}$, block header $I_H$, call depth $I_e$, and state modification permission bit $I_w$.

The accrued transaction substate $A$ is state accrued during a transaction's execution and is used to update EVM state immediately post-transaction. 
This is defined as a tuple $A \coloneqq (A_s, \substateLog, A_{\bm{t}}, A_r, A_{\bm{a}}, A_{\bm{K}})$, containing the self-destruct set $A_s$, series of logs emitted $\substateLog$, set of touched accounts $A_{\bm{t}}$, refund balance $A_r$, set of accessed account addresses $A_{\bm{a}}$ and set of accessed storage keys $A_{\bm{k}}$.


Finally, the 4-tuple of the above-defined state $(\worldState, \machineState, I, A)$ comprises the complete EVM state that can be read from or written to by a bytecode instruction.

\subsubsection{L2 execution semantics}\label{sec:oss}
We instantiate the L2 execution trace as described in \Cref{fig:state-transition} for the L2 execution state. 
We define the \textit{one-step state} (OSS), which encapsulates the state between execution trace steps as an authenticated data structure (ADS)~\cite{tamassia2003authenticated,miller2014authenticated}.
The OSS has three forms: \textit{intra-transaction state}, \textit{inter-transaction state}, and \textit{block state}.
We define the commitment of an OSS as $\HOSS$.

\paragraph{Intra-transaction state}
The \RollupVM \textit{intra-transaction state} $\sInit$ represents states between instruction execution.
It is directly constructed from the full EVM execution state $(\worldState, \machineState, A, I)$.
It contains every state field modifiable by an EVM opcode---including gas, stack, memory and world state. 

The commitment of the intra-transaction state does not directly hash the contents of fields that have inner structures (referred as \textit{components}), such as the stack, memory, and world state; instead, these components are also encapsulated using suitable ADSs (e.g. Merkle tree for the memory), and $\HOSS$ hashes on the commitments of these components.

This allows us to create the \textit{state proof} (described in \Cref{sec:prelim_proofs}) without having to reveal the full contents of every component.
A separate proof is submitted 
to prove the validity of the state transition in the component.
Since opcodes do not use every component (e.g. \texttt{ADD} doesn't access or modify the memory or the world state), the OSP provides only the necessary proofs. 

\paragraph{Inter-transaction state}
To encode EVM behavior that takes place between the consecutive execution of transactions, we define a special type of one-step state, the \textit{inter-transaction state} $\sInter$.
The inter-transaction state lies between the execution of two transactions, representing the finalized state after the execution of the first. 

\paragraph{Block state}
Similarly, to encode EVM behavior that takes place between the consecutive execution of blocks, we define another special type of one-step state, the \textit{block state} $\sBlock$.
The block state lies between two blocks, representing the finalized state of the first. 

During normal execution, commitments to the block state alone are computed and posted as \claims. See \Cref{app:oss} for a full state and commitment definition.
\subsection{Motivating example: \texttt{EXTCODECOPY}}

We begin with an example: verification of the \texttt{EXTCODECOPY} opcode, to illustrate the verification procedure.
\texttt{EXTCODECOPY} copies a bytecode segment from a contract other than the current execution environment, to memory.
The designated contract address and copying range are provided at the top of the stack.

The prover first reveals the current EVM intra-transaction state to the verifier using a \textit{state} proof.
The verifier checks if the revealed state is consistent with the commitment that both validators agreed upon during the interactive game.

The prover provides \textit{stack}, \textit{memory-write}, \textit{account-read}, \textit{code}, and \textit{opcode} proofs that partially reveal each corresponding EVM \textit{component} state so that the verifier can compute the state transition. 
\zhe{cross-ref to sec4.4?}
The verifier first checks if the stack proof is consistent with the stack commitment inside the state proof, and obtains the parameters of the \texttt{EXTCODECOPY} opcode, as well as the new \textit{stack hash} after these parameters are popped.
Given this proof, the verifier is able to compute the required gas for instruction execution and verify the validity of its parameters.
If a parameter is invalid or the gas charged exceeds gas available (revealed in the state proof), the verifier will execute the revert process, for which the prover is expected to provide the necessary proofs to prove the revert.

If validation is successful, the verifier then obtains the \textit{code hash} from the account-read proof, and the bytecode segment to be copied from the code proof.
Given the revealed bytecode segment, the verifier updates the new \textit{memory root} using the memory-write proof, which provides Merkle proofs of the designated memory writing range.
The verifier also updates the opcode field inside the state proof using the opcode proof to finalize the one-step \texttt{EXTCODECOPY} execution, as well as incrementing the program counter, charging gas, etc. 





\begin{algorithm}[t]
\caption{\new{One-step proof verification procedure}}
\label{algo:verification}
\begin{algorithmic}
\Procedure{Verify}{$\pi, \hInit, \hFinal$}
    \State assert $\hInit \stackrel{?}{=} \HOSS(\pi_{\sInit})$
    \State \textsc{VerifyData}($\pi$)
    \State $\sFinal \gets$ \textsc{Emulate}($\pi$)
    \State assert $\hFinal \stackrel{?}{=} \HOSS(\sFinal)$
\EndProcedure
\Procedure{VerifyData}{$\pi$}
    \State $c_{\texttt{actual}} \gets$ read on-chain commitment for $ \pi_{\texttt{data\_idx}}$
    \State assert $c_{\texttt{actual}} \stackrel{?}{=}$ \textsc{CommitData}($\pi_{\data}$)
\EndProcedure
\Procedure{Emulate}{$\pi$}
    \If{next step is a consensus validation step}
        \State $\sFinal \gets \textsc{EmulateConsensusStep}(\pi_\sInit, \pi_\data)$
    \Else
        \For{\textbf{each} component proof $\pi_c$ in $\pi_C$}
            \State $h_c \gets$ corresponding commitment in $\pi_\sInit$
            \State authenticate $\pi_c$ against $h_c$
        \EndFor
        \State $\sFinal \gets \textsc{EmulateExecStep}(\pi_\sInit, \pi_\data, \pi_C)$
    \EndIf
    \State \Return $\sFinal$
\EndProcedure
\end{algorithmic}
\end{algorithm}

\subsection{Proof system}
We describe the construction of a simple one-step proof system.
\ujval{frame as authenticated data structure}
At a high level, the one-step proof system provides a verification scheme to support the execution of a single trace step as an update operation\zhe{reference ADS for terminology?} of the OSS.
The prover attempts to update the OSS---whose commitment is known to the verifier---by executing one trace step, and convinces the verifier of the correctness of the execution through a one-step proof.

\begin{figure}[t]
  \centering
  \includegraphics[width=\linewidth]{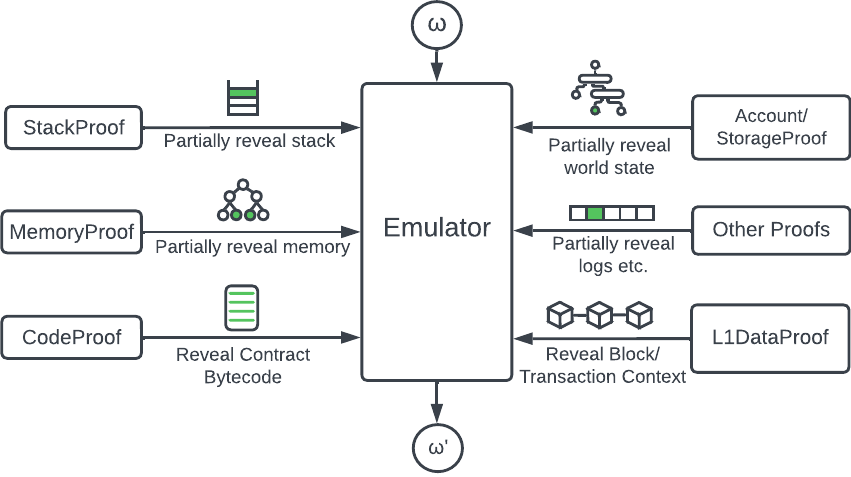}
  \caption{\textbf{Emulation.} An illustration of the flow of inputs and outputs within the one-step EVM emulator.}
  \captionspace{}
  \label{fig:verification}
\end{figure}

Formally, the one-step proof system $(P, V)$ consists of a prover $P(\sInit, \sFinal) = \pi$ and verifier $V(\pi, \hInit, \hFinal) = v \ | \ v \in \{0,1\}$, where $\hInit \coloneqq \HOSS(\sInit)$, $\hFinal \coloneqq \HOSS(\sFinal)$.
$\pi$ is a witness to a stateless \textit{emulator} (as outlined in \Cref{fig:verification}), which outputs the image of the L2 state-transition function if it were to be executed on the pre-image of $h$.
$\pi$ is a one-step proof of the transition $\sInit \to \sFinal$.
Specifically, $\pi$ authenticates $\sInit$, and establishes (1) that a single step results in $\sInit \to \sFinal$; and (2) $\sFinal$ hashes to the claimed $\hFinal$. 
$V$ outputs \ACCEPT if and only if $\pi$ proves the transition $\sInit \to \sFinal$, or \REJECT otherwise.
This procedure is captured in \Cref{algo:verification}.


The types of sub-proofs that $\pi$ consists of depend on the type of the state transition (for state transitions between intra-transaction states, also the opcode of the current instruction) and consequently which \RollupVM data structures are read from or written to. 
We describe each sub-proof below. 

\subsubsection{Common proofs}\label{sec:prelim_proofs}

We provide the following proofs for \textit{all} instructions in state transitions between intra-transaction states.
First, a \textit{state} proof is simply the pre-image of the OSS hash; this is provided for the initial state.
Second, an \textit{opcode} proof is the pre-image of the code hash, and authenticates the next instruction to be executed (at offset $\pcMS'$).
Third, an \textit{L1 data} proof $\pi_{\data}$ authenticates the execution environment information $I$ (c.f. \Cref{sec:evm}), including for example, the sender and recipient of the executing transaction, as well as the block context associated with the transaction (such as the timestamp and block number).
During normal operation (at sequencing-time), only a commitment to the execution environment information is saved on L1. 
Therefore, the prover is required to provide $\pi_{\data}$ to the verifier, which contains the calldata of transactions that were posted on L1.
The verifier can verify the correctness of $\pi_{\data}$ against the saved commitment and derive $I$ from $\pi_{\data}$.
We note that in a forthcoming upgrade, Ethereum will provide a special transaction primitive that automatically computes a KZG commitment over temporarily-stored data \cite{eip4844}. This can be used to verify data availability directly.

\subsubsection{Stack}\label{sec:stack}
The \textit{stack} proof attests to the validity of the state transition of the EVM stack $\stackMS \to \stackMSPrime$. 
We define the stack hash $\HStack$ as a simple hash-chain over the elements of the EVM stack $\stackMS$, from the bottom of the stack to the top.
This allows the prover to selectively reveal the top elements of the stack, read or popped by an EVM instruction, with only a single additional hash.
Specifically, the stack proof is the tuple $(h_0, \sPopped)$, where $h_0$ is the stack hash after elements are popped from the stack by the instruction, and $\sPopped$ is the subsequence of the elements popped from the stack.
The verifier verifies the correctness of $\sPopped$ by chain-hashing $\sPopped$ onto $h_0$ and comparing it against the stack hash committed to in $\sInit$.
To emulate a stack-push, the verifier can obtain the stack hash of $\stackMSPrime$ by chain-hashing $p'$ onto $h_0$, where $p'$ is the sequence of elements to be pushed to the stack after pops.

\subsubsection{Memory}\label{sec:memory}

Since the EVM's memory $\memMS$ is a form of virtual RAM, we must utilize a commitment that supports efficient random-access partial reveals and updates. 
An obvious choice is to use a Merkle tree over the \RollupVM memory space.
$\HOSS$ commits to a Merkle tree root computed over the entire contents of the \RollupVM memory space $\memMS$, where each leaf $i$ is $\KECCAK(\memMS[32i, 32i + 32])$.

The \textit{memory-read} proof $(i, c, p, r)$ authenticates the byte array $c$ starting at offset $i$ of $\memMS$, with respect to memory Merkle root $r$. 
The Merkle proof $p$ can be either a single-proof or a compact multi-proof, depending on the length of $c$.
A compact Merkle multi-proof \cite{ramabaja2020compact} authenticates multiple Merkle tree leaves succinctly, by aggregating Merkle proofs for tree leaves and de-duplicating the tree nodes in the proof.

Likewise, the \textit{memory-write} proof attests to the validity of a write to memory.
That is, it proves the validity of the new content at the location written to \textit{and} that the contents at no other location were changed.
This proof is of the form $(i, c, c', p, r, r')$, where $r'$ is the new Merkle root after the byte array $c'$ overwrites $c$ at the contiguous location starting at offset $i$ of $\memMS$.
The verifier first verifies $(i, c, p, r)$ as performed in the memory-read proof to verify $p$ is a correct proof of the consistency of $c$ in $r$.
It then performs memory writes based on $(i, c', p)$ to calculate the new Merkle root and compares with $r'$, the new memory Merkle root provided in the proof.
This proves that only the memory starting at location $i$ was modified.

The memory-read/write proofs are also applicable to the calldata $I_{\bm{d}}$ and return data $\returnMS$ of EVM since they act as read-only memory during transaction execution.

\paragraph{Unbounded memory access}
Certain EVM opcodes, including \texttt{KECCAK256} and those related to memory-copying, allow unbounded access to the EVM memory (subject to block gas limits).
Since the memory-read/write proof contains the accessed memory content, the proof size and therefore verification cost of such opcodes may be unbounded.
To prevent verification costs from exceeding the L1 block gas limit, we introduce a \textit{sub-step proof} allowing these opcodes to be treated as multiple sub-steps, each operating on a constant-size memory chunk.
As a concrete example, based on Keccak-256 hash function specification \cite{bertoni2013keccak}, the \texttt{KECCAK256} opcode is divided into three sub-step phases: 
\texttt{init}, which initializes a Keccak-256 hash state; 
\texttt{absorb}, which reads a constant size of the EVM memory and updates the hash state by emulating the Keccak-256 block permutation function in solidity;
and \texttt{squeeze}, which calculates the hash from the hash state and pushes the result back onto the EVM stack.

\subsubsection{World state}
\label{sec:world_state_reads}

\paragraph{Reads}
Recall that the account object and its storage are both encoded as Merkle Patricia trees (MPTs).
The \textit{account-read} proof authenticates the account state $\worldState[a]$ associated with address $a$, as read by an instruction. 
The MPT proof can be used to compute and verify the root as well as provide information corresponding to each account field 
(\accNonce, \accBalance, \storageRoot, \codeHash). 
Likewise, the \textit{storage-read} proof authenticates a value read from a storage slot in the MPT with root $\worldState[a]_s$ (assuming the consistency of $\worldState[a]$), as read by an instruction. 
It includes the content of the storage slot, with its MPT proof.

The \textit{code} proof similarly authenticates a contiguous sequence of instructions in an account's bytecode.
Similarly as in the case of the opcode proof, the code proof naively consists of the entire bytecode.
The verifier can verify the consistency of the bytecode by computing the code hash from the code proof and comparing it to the code hash stored in the contract account.

\paragraph{Writes}
A nice property of MPT proofs is that it allows the verifier to perform a single write (including update, insertion, and deletion) on the proven path and recalculate the trie root after the write.
This is possible because the MPT proof simply comprises all the relevant trie nodes along the proven path.
Therefore, the verifier can derive the trie root after the account write with a \textit{account-read} proof and the updated account, which is essentially the \textit{account-write} proof.
Similarly, the \textit{storage-write} proof includes a storage-read proof for the content during initial state $\worldState$, the new content to be written to the storage slot, and an account-read proof of the account where the storage belongs.

\subsubsection{Inter-transaction, block and consensus}\label{sec:inter_txn_proof}

The \textit{inter-transaction} proofs attest to the validity of the transaction initiation and finalization steps.
Depending on the type of transaction, the inter-transaction proof may include different types of proofs and additional information.
For example, the \textit{transaction initiation} proof always includes an \textit{account-read} proof of the sender and the contract to be executed, if the transaction is not regular.
Similarly, an \textit{account-write} proof is required for the \textit{transaction finalization} proof to authenticate the remaining gas refund, and code changes if the transaction is a contract creation.

The \textit{block} proof attests to the validity of the block initiation and finalization steps.
The \textit{block-initiation} proof simply reveals the field of the block state to be proven because there is no state change in the block initiation step.
The \textit{block-finalization} proof reveals the necessary state, such as the transaction trie root and receipt trie root for the verifier to reconstruct the block header.
It also provides the Merkle proof of the block hash tree, so that the verifier can verify the block hash update.

The \textit{consensus} proof authenticates the transaction data posted on L1 according to rollup-specific consensus rules.
Since transaction data is posted in batches, each batch only requires one step of consensus proof verification before the execution of any block in the batch to ensure the validity of the entire batch.
In its simplest form, the consensus proof enables direct witness checking: it authenticates transaction data, allowing the verifier to check if it is correctly encoded.

\subsubsection{Other components}\label{sec:other_proof}

There are some additional types of components inside the intra-transaction state, most of which are used for proving specific opcodes.
For example, for \texttt{LOG} opcodes, we accumulate all the logs emitted using a hash chain for the receipt calculation; for \texttt{SELFDESTRUCT} opcode we accumulate all the accounts that are self-destructed for transaction finalization; for \texttt{BLOCKHASH} opcode we use the Merkle proof of the block hash tree to authenticates block hashes of previous blocks.

\section{Implementation}
\label{sec:impl}

\begin{figure}
  \centering
  \includegraphics[width=\linewidth]{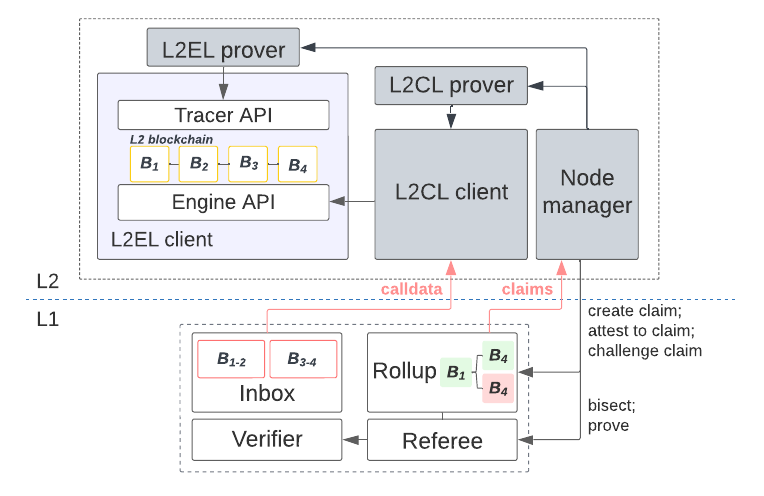}
  \caption{\textbf{\System architecture.}
  The L2 client software includes a modified Ethereum EL client, and custom CL client. 
  Gray arrows represent control flow and pink arrows represent data flow.
  Validators read, decode and apply batched transactions to a local EVM instance; 
  they also periodically post, attest to and/or dispute claims.
  }
  \captionspace{}
  \label{fig:arch}
\end{figure}

In this section, we describe the implementation of a new ORU, \textit{\System}.
\System's software stack consists of an L2 client software---including both an L2EL and L2CL client, as well as an auxiliary node manager; 
and an L1 bridge, implemented as a set of contracts. 

We focus NVP efforts on the most complex component in the stack, the L2EL client, which opportunistically reuses an Ethereum EL client for block building.
We have adapted two distinct Ethereum EL clients to support L2 operation: Geth \cite{geth-github} and Erigon \cite{erigon-github}. 
In this section, the former is referred to as \Client{}, while the latter as \ClientErigon{}.
While a fully trust-minimized system requires applying NVP to the L2CL client and prover, we leave this to future work.

The node manager, responsible for orchestrating validation, proving and batch submission (as a sequencer), is implemented as an external module, interfacing with Ethereum through standard JSON-RPC. 
The L1 batch submission, L2 chain derivation and L1 \claim creation are functionalities that operate at independent intervals. 
Claims can be posted at lower frequency relative to batch sequencing; 
therefore, a \claim may represent a state transition larger than a single sequenced batch.
Similarly, a batch may contain the transactions and metadata corresponding to several blocks.




\subsection{L2 client}

\subsubsection{L2CL client}
A validator's L2CL client is primarily responsible for orchestrating derivation of the L2 chain state as a function of L2 transaction date disseminated on L1. 
It monitors the L1 chain for new transactions to the inbox contract, indicating new transaction data has been made available. 
It then reads the calldata, containing encoded batches of transactions, along with metadata necessary to recompute the chain state deterministically (e.g. the associated L2 block numbers and timestamps). 
The metadata is validated and subsequently used to reconstruct execution payloads, containing a list of block payloads, for local chain insertion. 
These payloads are delivered to the local L2EL client using Engine API, following the approach introduced by Bedrock \cite{optimism-bedrock}.
\System's L2CL client is a fork of a Rust-based Optimism client \cite{magi}---modified to support a simpler consensus-level semantics.

\subsubsection{L2EL client}
The L2EL client maintains the local blockchain. 
It applies fork-choice update requests from the local (authorized) L2CL client, marking blocks as \texttt{safe} and \texttt{finalized} once they appear on-chain, and are finalized on Ethereum respectively.
The L2EL client is also responsible for building \texttt{unsafe} blocks from payload-building requests, using transactions from the local transaction pool if allowed to, and from the request if explicitly provided. 

As mentioned previously, \System opportunistically leverages two L2EL clients, adapted from the Ethereum ecosystem---Geth and Erigon. 
\System is secure as long as one L2 client software stack is implemented correctly.

\paragraph{Geth}
We adapt Geth to L2 and implement all necessary functionality for both normal operation and dispute resolution. 
This includes support for the prover and deterministic execution via Engine API. 
To support the proof generation, 99 lines of code are introduced to Geth's StateDB implementation, extending the \texttt{EVMLogger} API.

\paragraph{Erigon}
Erigon\cite{erigon-github} is a deeply modified Geth fork, focusing on achieving high performance.
It has undergone extensive modifications and has diverged significantly from the original Geth codebase since July 2020, making it a non-trivially distinct Ethereum client implementation.
Erigon uses staged sync \cite{erigon-github-staged-sync} for local blockchain construction, and relies on multiple caching layers in their EVM implementation.

Therefore, wrappers over some Erigon components, such as StateDB, are necessary to ensure that Erigon implements all interfaces required by the proof module (e.g. copying StateDB results in a deep-copy).
We adapt Erigon to support OSP generation, modifying and adding only \ErigonProverLoC lines-of-code. 
A shim is responsible for implementing the necessary interfaces between the prover (same as that used for Geth) and \ClientErigon{}.
The Erigon IFP shim consists of \textasciitilde1.3k lines of Golang code. 

We note that \ClientErigon{} is an incomplete prototype intended to demonstrate NVP capability. 
Missing functionality includes support for deterministic payload building via Engine API and fee semantics.
However, there are no fundamental barriers to implementing this set of functionality.


\subsubsection{Node manager}
The node manager is responsible for creating, attesting to and disputing claims, as a function of the local L2 chain state it derives \textit{and} the L1 claim chain contract state.
On observing a claim which conflicts with one that it has already posted and validated, the validator can invoke the dispute initiation mechanism in the bridge contract against the claim creator, or any other validator that has attested to it. 
It then participates in an IFP against another validator and invokes the prover to produce a one-step proof at the end of the dispute.
This follows the simple protocol introduced by Aribtrum \cite{kalodner2018arbitrum}; however, the system and proof scheme is also generally compatible with newer protocols \cite{nehab2022permissionless, alvarez2023bold}.


\subsubsection{Prover}
We implemented the \textit{prover} (illustrated in \Cref{fig:arch}), which exposes a set of RPC APIs to validators for commitment and OSP generation. 
During normal execution, the prover is able to produce claims by constructing the inter-transaction states directly from transaction data or receipts.

When a dispute begins, the validator re-executes the transaction that is ultimately disputed, and interacts with \RollupVM through an \textit{extended} \texttt{EVMLogger} API, which records sufficient \RollupVM state information in each execution step.
The prover relies on the API to obtain internal \RollupVM states that are not exposed by standard Ethereum APIs.
For instance, the extended API includes a \text{copy} function, enabling the prover to obtain a \textit{deep-copy} of a certain \RollupVM state.
This recorded \RollupVM state information at the disputed execution step is then used to construct one-step states for intermediate commitments and to generate the final OSP.
The prover is implemented in \textasciitilde5k lines of Golang code. 

\subsection{L1 bridge}

The L1 bridge includes the referee and OSP verifier for the scheme outlined in \Cref{sec:evmnative}. 
The L1 bridge contracts are implemented in Solidity \texttt{0.8.4}, and follow the design of Arbitrum's \cite{kalodner2018arbitrum}---the key difference being in the OSP verifier. 
The bridge functionality is split across:
(1) a sequencer inbox, where L2 transaction data is made available within the calldata on the L1 blockchain;
(2) a claim chain, where claims are created, attested to and disputed; 
(3) a referee, which is deployed by the claim chain contract upon dispute initiation, to referee an interactive game; and 
(4) the OSP verifier, called in the final round. 

The OSP verifier comprises of verifier contracts and library contracts.
To avoid exceeding the maximum contract size limit, we split the verifier into a \texttt{VerifierEntry} contract and a set of \texttt{Verifier} contracts.
The former dispatches OSP verification requests to different subproof-specific verifiers.
The \texttt{Verifier} contract set consists of 8 different verifiers, each of which implements verification of several types of one-step proofs.
The library contracts provide common utilities for all verifiers, including one-step proof definition and decoding, verification context, EVM gas metering, and EVM consensus parameters.
The verifier is implemented in \textasciitilde4k lines of Solidity code. 
\begin{figure}[t]
    \centering
    \begin{subfigure}[t]{0.49\linewidth}
    \centering
    \includegraphics[width=\linewidth]{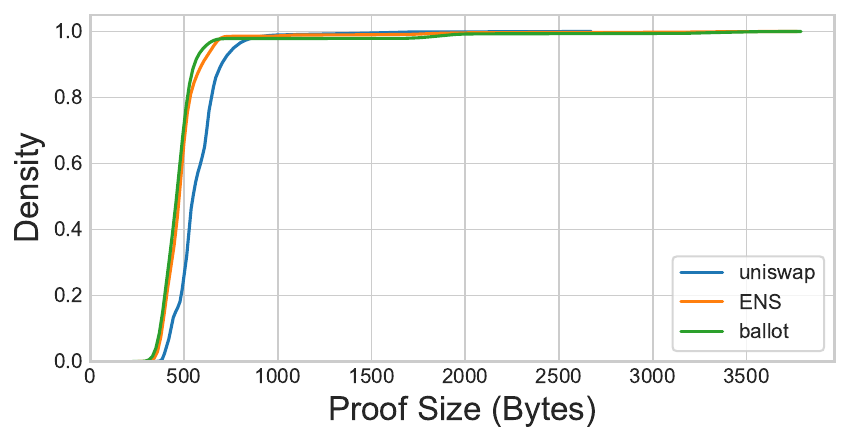}
        \subcaption{Proof size CDF}
    \end{subfigure}
    \begin{subfigure}[t]{0.49\linewidth}
    \centering
    \includegraphics[width=\linewidth]{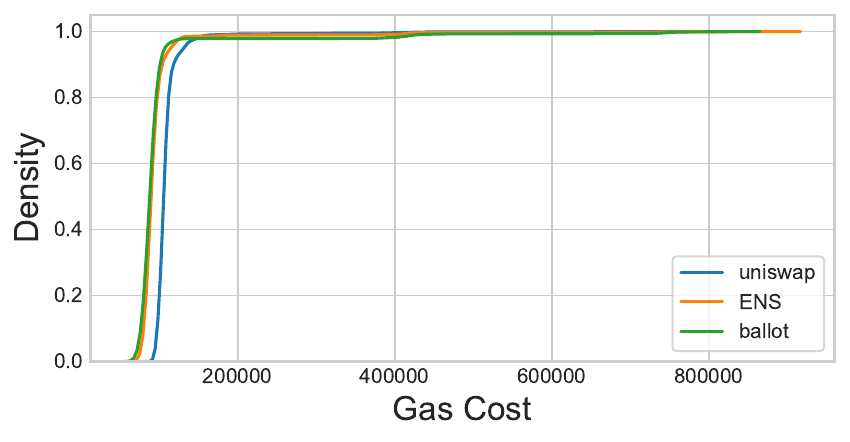}
        \subcaption{Gas cost CDF}
    \end{subfigure}
\caption{\textbf{Proof sizes and verification gas costs for transactions of 3 evaluated applications (sans contract effects).} 
    (a) shows the distribution of one-step proof sizes.
    (b) shows the distribution of verification gas costs.}
  \captionspace{}
\label{fig:exp-cdf}
\end{figure}


\section{Evaluation}

\begin{figure*}
    \centering
    \includegraphics[width=\linewidth]{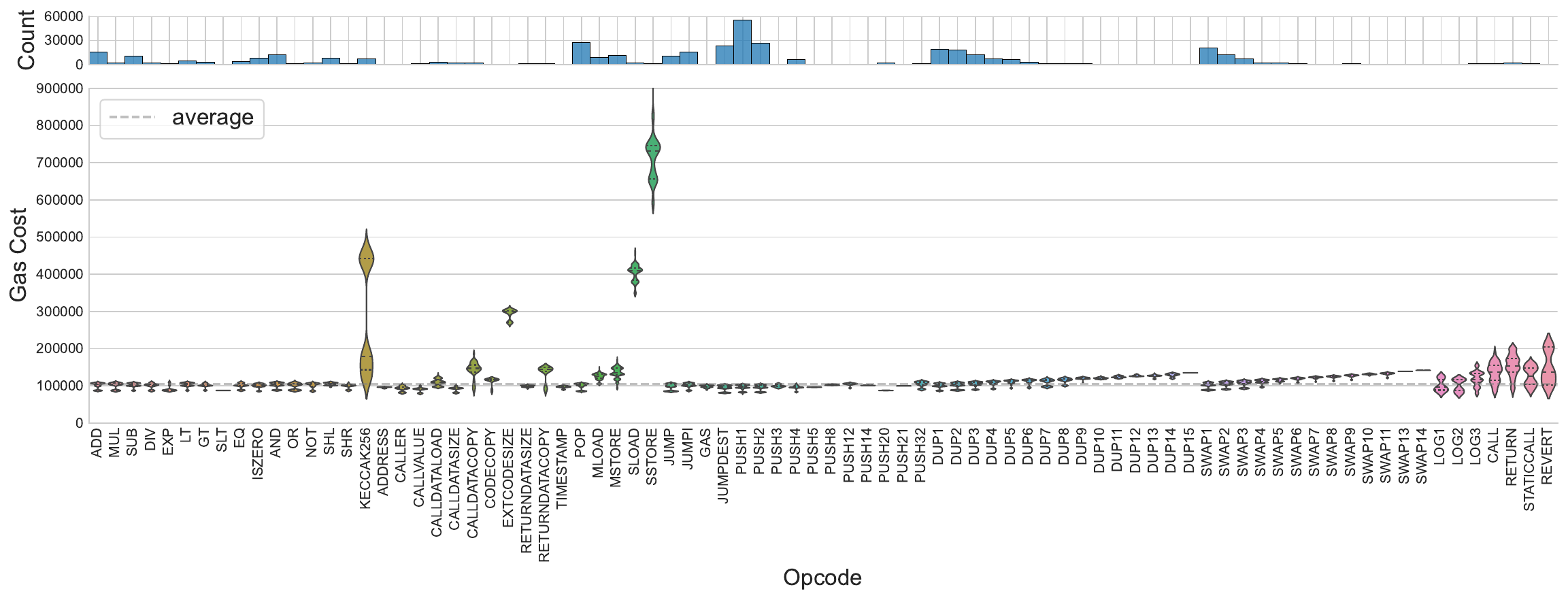}
\caption{\textbf{Verification gas costs for evaluated transactions (sans contract effects).} 
    Illustrates gas cost distribution on a per-opcode basis, along with the distribution of opcodes in the contract.
}
  \captionspace{}
\label{fig:exp-gascost}
\end{figure*}

In this section, we aim to show that our system's dispute resolution performance is practical. 
We leave performance improvements to future work and discuss low-hanging fruit in \Cref{sec:future}.
All experiments are performed on a desktop with an Intel i9-13900K @ 5.80 GHz processor, 64 GB memory, and Ubuntu Windows Subsystem for Linux.
To study the performance characteristics of our OSP in real world conditions, we carry out disputes on state transitions resulting from interactions with 3 popular applications, with setups outlined as follows.

\paragraph{Uniswap V2}
Uniswap V2 is a popular decentralized automated market maker.
We deployed the Uniswap V2 contract \cite{uniswap-github}, along with 4 ERC-20 token contracts, on \System.
We then generated and executed 100 transactions, each of which is a call to one of the following Uniswap contract functions (sampled uniformly at random) with random parameters: \texttt{Add\-Liquidity}, \texttt{Remove\-Liquidity}, \texttt{Swap\-Tokens\-For\-Exact\-Tokens}, \texttt{Swap\-Exact\-Tokens\-For\-Tokens}.
These functions will interact with the following contracts (with bytecode size in parenthesis): \texttt{ERC20} (2.1KB), \texttt{Uniswap\-V2\-Pair} (8.6KB), and \texttt{Uniswap\-V2\-Router\-02} (17.5KB).

\paragraph{Ethereum Name Service}
Ethereum Name Service (ENS) is the most popular naming service on the Ethereum blockchain, and one of the largest NFT applications.
We deployed the ENS registry contract \cite{ens-github} on \System.
In this experiment, we generate and execute 100 random ENS registration transactions, each of which interacts with the \texttt{ENSRegistry} contract (4.2KB).

\paragraph{Ballot}
On-chain voting is also a popular blockchain application used by numerous DAOs for governance.
We deployed a basic voting contract called ballot \cite{ballot-github} on \System and generated 100 random voting or delegation transactions for evaluation.
These transactions interact with the \texttt{Ballot} contract (3.8KB).


For evaluation purposes only, we construct one-step states and generate one-step proofs for \textit{all} executed state transitions (during normal execution in a real deployment, we do neither). 
In total, we execute 300 transactions, resulting in a total of \textasciitilde400k steps. 
We construct one-step states for each step for the purpose of this experiment.
We then evaluate one-step proof generation and verification on each executed state transition in next sections.
\Cref{fig:exp-cdf} illustrates the overall proof size and verification gas cost distributions.

As mentioned in \Cref{sec:prelim_proofs}, we naively include the entire contract bytecode in the proof, to prove consistency of the bytecode.
Therefore, to better illustrate the actual properties of the one-step proof evaluated, Figures 7 and 8 do not include the effects of contract bytecode (however, the associated text in this Section does).
We note that \System's approach is practical even if the entire contract bytecode is included in the proof---as shown in the next subsections. 





\subsection{One-step proof size}
We measure the latency of generating a one-step proof (across \textasciitilde400k steps) on \Client{} to be \textasciitilde0.739ms on average. 
Proof generation therefore has negligible latency.

The average size of the one-step proofs generated---without contract bytecode included---is 558B (min. 323B, max. 3684B).
The proof sizes \textit{with} contract bytecode included for opcode and code proofs are an average of 11.6KB (min. 2.63KB and max. 29.6KB).
This size depends largely on the contract in which the step occurs.
As a baseline, Arbitrum claims that the average size of an AVM one-step proof is \textasciitilde200B, with a maximum of \textasciitilde500B.
However, we expect the proof sizes in both Nitro and Cannon to be significantly larger than that of the AVM, since they generate one-step proofs on lower-level ISAs, which use a flat memory model like that of the EVM (no experimental results have been made public yet by either project).

The EVM is not designed to generate succinct one-step proofs.
However, given the rarity of disputes, we argue that the OSP sizes are practical. 

\subsection{One-step proof verification cost}
The average gas cost for the verification of the one-step proofs generated---again, ignoring contract size effects---is 109k gas (min. 78k, max. 897k). 
For reference, a typical Uniswap v2 swap transaction on Ethereum consumes \textasciitilde170k gas.
This is 2-3 orders of magnitude under the Ethereum block limit. 
\Cref{fig:exp-gascost} provides a distribution of verification gas cost on a per-opcode basis.

When contract bytecode is included in one-step proof for either opcode proof or code proof, the average gas cost is 629k gas (min. 155k, max. 1,874k).
This increase in gas cost includes the cost of including the contract bytecode in calldata, copying it into memory, and hashing to verify its consistency.
Given that contracts deployed on Ethereum cannot exceed the size limit of 24KB, we estimate that the maximum gas cost in the worst case will not exceed 3,000k gas (i.e. only 10\%{} of the Ethereum block gas limit).
Thus, the worst case verification cost is still practical.

Aside from the gas cost introduced by the contract bytecode inclusion, a significant factor in proof verification gas cost is MPT proof verification for opcodes that access the world state, as shown in \Cref{fig:exp-gascost}.
\texttt{SSTORE} requires four MPT proofs, including two for the account and storage (one for the storage slot of the committed state for gas charges and one for the storage slot write).
For opcodes that may access unbounded memory, the gas costs of verifying different sub-steps are grouped together.
For \texttt{KECCAK256} in particular, there are significant differences in the gas costs associated with verifying each of its three types of sub-steps---giving the appearance of a bimodal distribution.


\subsection{Erigon proof generation}
We also evaluated the \ClientErigon{} prover on the randomly generated 300 transactions.
The \ClientErigon{} prover successfully generated the same proofs (non-determinism such as block timestamps aside) as \Client{}, which were able to pass verification.
We measure the latency of generating a one-step proof on \ClientErigon{} to be \textasciitilde6.5ms on average.
While 10x slower than \Client{} because of the adapter efficiency, the proof generation latency is also negligible.

\section{Discussion}

\subsection{Related Work}
\label{sec:related}

Existing rollups can be categorized as either \textit{optimistic rollups} or \textit{validity rollups} (also colloquially referred to as zk-rollups). 
We now touch on the relevance of our work in the context of the latter.
In a validity rollup, a state update posted on-chain must be accompanied with a validity proof (for example, a SNARK) that convinces the L1 verifier of the correctness of the computation. 
The security of a validity rollup relies on the soundness of the SNARK verifier.
The TCB therefore contains a complex SNARK compiler and proof verifier that use heavy cryptographic machinery. 
Not only can this be difficult to understand and audit, there is a non-trivial risk of encountering soundness or circuit-level semantic bugs \cite{dahlgren2022circomspect, block2023fiat, pailoor2023automated}.
Incidentally, this machinery also poses a challenge for popular formal methods tooling, such as SMT solvers.
While there has been some progress towards the formal verification of simple SNARK constructions \cite{bailey2023formalizing}, proofs of soundness for more complex constructions, such as those used by zkEVM systems, have thus far been out of reach. 

\subsection{Future Work}
\label{sec:future}

\paragraph{Formal verification}
We plan to formally verify our L1 verifier against an existing executable Ethereum formal specification \cite{cassez2023formal, hildenbrandt2018kevm} to further reduce the TCB size.
Specifically, the objective is to verify the correctness of the EVM bytecode compiled from the Solidity contracts, eliminating the compiler from the TCB in the process.

However, directly verifying against the specification at the EVM bytecode level is beyond the capabilities of current tooling. 
Fortunately, Solidity compilers operate on a semantically-rich intermediate representation (IR), Yul.
Therefore, to achieve modular and incremental verification, one approach is to leverage Dafny to 
(1) verify the L1 verifier against the specification \textit{at the Yul-level}; 
and (2) establish a refinement relation between the verifier IR and bytecode.
This removes trust in the compiler altogether.

\paragraph{Lower-cost proof verification}
The proof system relies on naive instruction emulation to introduce minimal cryptographic assumptions and facilitate the application formal methods.
This comes at the price of verification performance and cost.
However, this can be easily addressed through a two-tiered proof system design as follows, \textit{without} significantly modifying the TCB.

A more efficient proof system ($P_1$, $V_1$) (e.g. using a SNARK \cite{pse}) is chosen to serve as a default fast-path, while the simpler proof system ($P_2$, $V_2$) serves as a more trustworthy fallback, \textit{only} in case the verifier $V_1$ is differentially determined to be unsound off-chain.
This is made possible by simply adding an additional short time delay post-verification by $V_1$, in which a disagreeing proof may be submitted to $V_2$.
There is no additional on-chain cost, except when $V_1$ is unsound.

\paragraph{Lower-cost normal execution}
The majority of the cost to transact is a function of storing transaction batch data to L1.
We can therefore reduce L2 transaction fees by an order of magnitude through the compression of transaction batch data prior to sequencing. 
DEFLATE \cite{rfc1951} is used by other rollups \cite{optimism-bedrock} and works well in practice. Moreover, because DEFLATE is well-specified by its RFC and comes packaged in several compression libraries, it lends itself to $1$-NVP.

\section{Conclusion}

In this work, we examine the problem of designing and implementing a secure, trust-minimized optimistic rollup. 
We study the connection between IFP protocol design and NVP, and outline the properties of prior and existing ORU systems in this context.
We propose an approach that addresses the limitations of the state-of-the-art, and realize this approach in practice by building \System, an ORU that leverages opportunistic $1$-NVP to provide a secure, trust-minimizing scalability solution for Ethereum.

\ifauthor

\section*{Acknowledgement}
This material is in part based upon work supported by the Center for Responsible, Decentralized Intelligence at Berkeley (Berkeley RDI). 
Any opinions, findings, and conclusions or recommendations expressed in this material are those of the author(s) and do not necessarily reflect the views of these institutes.
We thank Patrick McCorry and Franck Cassez for their thoughtful feedback, which greatly improved the manuscript.
\fi

\bibliographystyle{unsrt}
\bibliography{refs}

\appendices
\crefalias{section}{appendix}
\begin{table*}[t]
  \centering {\small
  \begin{tabular}{|c|c|}
    \hline
    \textbf{State type} & \textbf{Structure} \\
    
    \hline
    \texttt{inter-block} & 
    \begin{math}
        \texttt{batch} \ || \ \texttt{blockNumber} \ || \ r(\worldState) \ || \
        \texttt{cumulativeGasUsed} \ || \ r(T_b)
    \end{math} \\ 

    \hline
    \texttt{inter-tx} &
    \makecell{
        \centering
       $\begin{gathered}
        \texttt{batch} \ || \ \texttt{blockNumber} \ || \ \texttt{transactionIdx}  \ || 
        \ r(\worldState) \ || \texttt{cumulativeGasUsed} \ || \\
         \texttt{blockGasUsed} \ || \ r(T_b) \ || \ r(T_t) \ || \ r(T_r)
    \end{gathered}$
    } \\

    \hline
    \texttt{intra-tx} &
    \makecell{
    $\begin{gathered}
            \texttt{batch} \ || \ \texttt{blockNumber} \ || \ \texttt{transactionIdx}  \ || I_e \ || \ \machineState_g \ || \ A_r \ || \ \HOSS(\texttt{lastDepthState}) \ || \ I_a \ || \ {\worldState[I_a]}_{c} \ || \\ 
            \ I_s \ || \ I_v \ || \ \texttt{callFlag} \ || \ \texttt{out} \ || \ \texttt{outSize} \ || \ \pcMS \ || \ \instructionEE \ || \ size(\stackMS) \ || 
            \  \HStack(\stackMS) \ || size(\memMS) \ ||  \\ 
            \ r(\memMS) \ || \ size(I_{\bm{d}}) \ || \ 
            r(I_{\bm{d}}) \ || \  size(\returnMS) \ || 
            \ r(\returnMS) \ || \ r(\worldState) \ ||  H(A_t) \ || \ r(\texttt{accessList}) \ || \ H(A_l)
        \end{gathered}$
    } \\

    \hline
  \end{tabular}
  }
    \caption{
    \textbf{One-step state serialization structure.\zhe{I added word serialization}} 
    Passed into a collision-resistant hash function (e.g. $\KECCAK$).
    $size(\cdot)$ represents the size of the input structure in bytes, and $r(\cdot)$ represents the root hash of a Merkle tree or a Merkle Patricia tree. 
    Note: some parameters are omitted when they are not needed or can be derived.
    }
\end{table*}


\section{One-step proof}

\subsection{One-step state definition}\label{app:oss} 
We elaborate on the structure of the \textit{one-step state} $\sInit$ and the structured commitment to it below (including notation defined in \cite{wood2014ethereum}):

\paragraph{Block state}
A block state $\sBlock$ has the following fields to completely describe the state of a finalized block:
\setlist[description]{font=\normalfont\tt\space}
\begin{description}
    \item[batch] The batch number of the current block.
    \item[blockNumber] The number of the current block.
    \item[$\worldState$]  The EVM world state after finalization of the current block represented in a \textit{trie}.
    \item[cumulativeGasUsed] Cumulative gas used by this block and all its ancestors.
    \item[$T_b$] a Merkle tree of 256 previous block hashes (including current block). The block hash of block $i$ is stored at the leaf of index $i \%{} 256$, given block $i$ is the current block or one of the 255 ancestors of the current block.
\end{description}


\paragraph{Inter-transaction state}
An inter-transaction state $\sInter$ has the following fields to completely describe the state between transactions:
\begin{description}
    \item[batch] The batch number of the current block.
    \item[blockNumber] The number of current block.
    \item[transactionIdx] The idx of the transaction right before this state.
    \item[$\worldState$] The EVM world state after finalization of the previous transaction, represented in a \textit{trie}.
    \item[cumulativeGasUsed] Cumulative gas used within the current block.
    \item[blockGasUsed] Cumulative gas used before the current block.
    \item[$T_b$] the Merkle tree of 256 previous block hashes (excluding current block).
    \item[$T_t$] the transaction list with all transactions before this state, represented in a \textit{trie}.
    \item[$T_r$] the receipt list with receipts of all transactions before this state, represented in a \textit{trie}.
\end{description}

\paragraph{Intra-transaction state}
An intra-transaction state $\sInit$ has the following fields to completely describe the EVM state during its execution:
\begin{description}
    \item[batch] The batch number of the current block.
    \item[blockNumber] The number of the block where the current transaction belongs.
    \item[transactionIdx] The index of the current transaction in the block.
    \item[$I_e$] The execution depth (i.e. how deep the call stack is) in the current point of execution. If the current executing contract is directly called by the transaction, the $I_e$ is 1.
    \item[$\machineState_g$] The gas available to the current call.
    \item[$A_r$] The gas to refund at the end of execution.
    \item[lastDepthState] The OSS of the caller at the time when the current contract is called without calling arguments on stack. If $I_e$ is 1, the \texttt{lastDepthState} is $\worldState_0$, the EVM checkpoint state before transaction execution.
    \item[$I_a$] The address of the current executing contract.
    \item[$I_s$] The address of the caller.
    \item[$I_v$] The value that passed along with the call in the current point of execution.
    \item[callFlag] The type of calling opcode is used when the current contract is called. 0 for \texttt{CALL}, 1 for \texttt{CALLCODE}, 2 for \texttt{DELEGATECALL}, 3 for \texttt{STATICCALL}, 4 for \texttt{CREATE}, 5 for \texttt{CREATE2}. If $I_e$ is 1, the \texttt{callFlag} is always 0 if the transaction is a contract call, or 4 if the transaction is a contract creation.
    \item[out] The starting offset of where the return data should be copied to the caller's memory when the current contract returns.
    \item[outSize] The size of the return data that should be copied to the caller's memory when the current contract returns.
    \item[$\pcMS$] The offset of the current executing opcode.
    \item[$\instructionEE$] The current executing opcode. 
    \item[${\worldState[I_a]}_{c}$] The \codeHash in the account state of the current executing contract.
    \item[$\stackMS$] The EVM execution stack in the current point of execution, with each element of 256 bits. The hash of the stack is defined of hash chain of the elements in the stack, starting with the bottom of the stack. The hash of an empty stack is zero hash.
    \item[$\memMS$] The EVM execution memory in the current point of execution as a byte array. The byte array is segmented in 256 bits to form cells. The last cell is padded with 0s if its length is less then 256 bits. Then cells are stored in a Merkle tree, the leaf node of which is in format of \texttt{offset} || \texttt{cell}, where \texttt{offset} is the offset of the cell.
    \item[$I_{\bm{d}}$] The input data to the contract being executed. It is built as a Merkle tree similar to $\memMS$.
    \item[$\returnMS$] The return data of the last returned call, empty if no contract is returned yet. It is built as a Merkle tree similar to $\memMS$.
    \item[$\worldState_0$] The World State of EVM before the transaction execution represented in a \textit{trie}.
    \item[$\worldState$] The World State of EVM in the current point of execution represented in a \textit{trie}.
    \item[$A_t$] the self-destructed set. In \System, the hash of the self-destructed set is defined as the hash chain of self-destructed contract addresses in order.
    \item[$A_l$] logs emitted during the transaction. The hash of the log series is defined as the hash chain of all logs emitted in order.
    \item[$T_b$] the same block hash tree as of the inter-transaction before the current transaction.
    \item[$T(A_a, A_K)$] the set of accessed account addresses and storage keys for changing gas metering behaviors. In \System, the access list is constructed as a Merkle Patricia trie where the key is the account address and the value is the storage keys accessed.
\end{description}

\newpage


\section{Meta-Review}

The following meta-review was prepared by the program committee for the 2024
IEEE Symposium on Security and Privacy (S\&P) as part of the review process as
detailed in the call for papers.

\subsection{Summary}
This paper presents Specular, an L2-native and interactive fraud proof optimistic rollup (ORU) framework and provides concrete instantiations for EVM-based systems. The main issue of current EVM-based ORUs that Specular addresses is that the rollup program binary and the on-chain verifier are too tightly coupled, which leads to their inability to support n-version programming, complex trusted computing base, and opaque upgrade processes.

\subsection{Scientific Contributions}
\begin{itemize}
\item Provides a Valuable Step Forward in an Established Field
\end{itemize}

\subsection{Reasons for Acceptance}
\begin{enumerate}
\item L2-native interactive fraud proof ORU is a novel solution to known problems in EVM-based ORUs, with the advantages of supporting diverse Ethereum clients and reducing the size of TCB.
\item The paper demonstrates careful and reasonable state commitment and proof engineering considerations.
\end{enumerate}

\end{document}